\documentclass[10pt]{iopart}
\usepackage{epsfig,graphicx}

\usepackage{amsfonts}

\newcommand{\V}{\mathcal{V}}
\newcommand{\F}{\mathcal{F}}
\newcommand{\G}{\mathcal{G}}
\newcommand{\setS}{\mathcal{S}}

\begin{document}

\title[Spin systems on hypercubic Bethe lattices]{Spin systems on hypercubic Bethe lattices: \\A Bethe-Peierls approach}%

\author{Alexander Mozeika$^{\dag}$ and Anthony CC Coolen$^{\dag\ddag}$}
\address{$\dag$ Department of Mathematics,  
King's College London, 
The Strand, 
London
WC2R 2LS, UK
\\
$\ddag$ London Institute for Mathematical Sciences, 35a South St, Mayfair, London W1K 2XF, UK }


\ead{alexander.mozeika@kcl.ac.uk, ton.coolen@kcl.ac.uk}



\begin{abstract}
We study spin systems on Bethe lattices constructed from $d$-dimensional hypercubes. Although these lattices are not tree-like, and therefore closer to real cubic lattices than Bethe lattices or regular random graphs, one can still use the Bethe-Peierls method to derive exact equations for the magnetization and other thermodynamic quantities.  We   compute phase diagrams for ferromagnetic Ising models on hypercubic Bethe lattices with dimension $d=2, 3,$ and $4$.  Our results are in good agreement with the results of the same models on $d$-dimensional cubic lattices, for low and high temperatures, and offer an improvement over the conventional Bethe lattice with connectivity $k=2d$. 
\end{abstract}



\section{Introduction\label{section:intro}}
Lattice spin systems are idealized mathematical models of magnetic materials.
In the absence of external disturbances such systems are in equilibrium, and governed by the Gibbs-Boltzmann distribution $P(s)=\rme^{-\beta E(s)}/Z$, where $E(s)$ is the energy (or Hamiltonian) of a micro-state $s\in \setS^N$, $\beta=1/k_BT$  is the inverse rescaled temperature, and  $Z=\sum_s\rme^{-\beta E(s)}$  is the partition function. If the set $\setS$  of individual spin states is continuous, the sum in $Z$ becomes an  integral. The averages of macroscopic functions of micro-states (observables), such as the total energy or the magnetization, can be obtained  by differentiation from the free energy $F=-T\log Z$, which is related to the internal energy $U=\langle E(s)\rangle=\sum_s P(s) E(s)$ and the Gibbs-Shannon entropy $S=-\sum_s P(s)\log P(s)$ via the thermodynamic relation $F=U-TS$. However, computing $F$ analytically for an interacting system of macroscopic size is difficult, and to date only few lattice spin models models have been solved exactly~\cite{Baxter1982}.

To circumvent the above problem one often approximates the true micro-state distribution $P(s)$ with a simpler alternative $P_0(s)$, which retains only some characteristics of the original model. This approximation can usually be interpreted as a deformation of the true topology of the lattice such that short loops are removed and the analytical computations of thermal averages become easier, 
in combination with a variational approach that utilises the inequality $\sum_s P_0(s)\log[P_0(s)/P(s)]\geq 0$. The variational mean-field (v-MF) approximation, see e.g.~\cite{Opper2001},  uses the probability distribution  $P_0(s)$ of a non-interacting system  in this  inequality. For $N\rightarrow\infty$ its   results are  equal to the results of exactly solvable  (ferromagnetic) spin systems on complete graphs~\cite{Opper2001}, but for $d$-dimensional lattices they are unreliable. It predicts incorrectly, for instance, a phase transition in  the $1$-dimensional Ising model.  However, its predictions for critical exponents are correct for ferromagnetic spin systems when $d>4$;  see~\cite{Heydenreich2008} for a unified proof of this result and references  to relevant earlier work .

In the Bethe-Peierls (BP) approximation  (also known as belief propagation in computer science~\cite{Yedidia2005}, or  the cavity method in the spin glass community~\cite{Mezard2001}) one replaces the original lattices by  tree-like graphs,  which enables a recursive computation of thermal averages. One such graph is the Bethe lattice~\cite{Bethe1935}, usually defined as the `central' part of an infinitely large Cayley tree~\cite{Baxter1982}.  Closely related to the Bethe lattice is the random regular graph (RRG), defined as a maximally random graph in which all vertices have the same degree~ \cite{Bender1978, Bollobas1980}. RRGs do have loops, but these are typically of  length $O(\log N)$ as $N\rightarrow\infty$, so RRGs are locally tree-like.  For ferromagnetic models,  Bethe lattices and  RRGs give the same results \cite{Johnston1998},  but  in antiferromagnetic and spin-glass models the loops in the RRGs generate frustration, and can not be ignored~\cite{Mezard2001}.  The BP approximation is more reliable than the 
MF approximation~\cite{Gujrati1995},  since it involves a less brutal deformation of the original lattice, and it is 
exact for ferromagnetic Ising models on locally tree-like random graphs~\cite{Montanari2010}; it is interesting that, despite the fact that they can be solved relatively easily, the behaviour of Ising models on trees is more complex than in $d$-dimensional lattices~\cite{Haggstrom1996}. Further improvements of the BP approximation were obtained by correcting the BP solution for rare loops~\cite{Montanari2005, Parisi2006}; the improved theory is exact for a Bethe lattice with exactly one loop~\cite{Montanari2005}.  

In this paper we study spin models in which not only the correct coordination numbers of $d$-dimensional cubic lattices are retained, but (unlike the v-MF, BP, and RRG approximations) also the statistics of short loops and many of their nestings.  The spins in our models occupy the vertices of 
  Bethe-type lattices constructed from  $d$-dimensional hypercubes, i.e. from the cells (squares, cubes, etc.) of  the conventional $d$-dimensional cubic lattice.  These hypercubic  Bethe lattices  can be seen as generalisations of Husimi lattices~\cite{Jurcisinova2012}, which are Bethe lattices constructed from loops such that no edge lies in more than one loop. We use the Bethe-Peierls method to derive equations for the average magnetization, the specific heat, and the internal energy per spin. From these we compute phase diagrams for the ferromagnetic Ising version of the model with $d=2,3,4$. Our phase diagrams are compared with Onsager's exact result for the $d=2$ square lattice, and with high- and low-temperature expansions and Monte Carlo (MC) simulation results for  cubic lattices with $d=3,4$.

\section{Spin systems on factor trees and the Bethe-Peierls method\label{section:BP}}

\begin{figure}[t]
\vspace*{-7mm}
\setlength{\unitlength}{1mm}
\begin{center}{
\hspace*{10mm}
\begin{picture}(100,80)
\put(0,0){\includegraphics[height=75\unitlength]{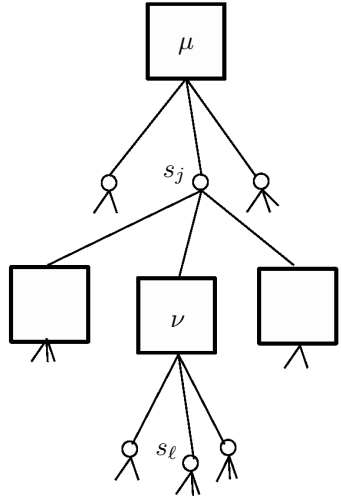}}
\put(36,64){$\mu$}
\put(34,47){$s_j$}
\put(35,27){$\nu$}
\put(33,10){$s_\ell$}
\end{picture}
}\end{center}
\vspace*{-6mm}
\caption{The interaction topology of a spin system on a factor-tree rooted at factor-node $\mu$. All spins are represented by circular `variable' nodes, and each term in the  energy  (\ref{def:E}) corresponds to a square `factor' node. A link between variable node $\ell$ and factor node $\nu$  implies that $s_\ell$ acts as an argument of $E_\nu(s)$.}
\label{figure:factor-tree} 
\end{figure}

We know that the interaction topology of any spin system of size $N$ can be represented by a a bipartite factor-graph $\G=(\V, \F)$ \cite{Mezard2009}, 
with $N=\vert\V\vert$ variable-nodes and  $M=\vert\F\vert$ factor-nodes, in which 
the micro-state energy takes the general form
\begin{eqnarray}
E(s)=\sum_{\nu\in \F} E_\nu(s).\label{def:E}
\end{eqnarray} 
Here we denote with  $s=\{s_i\!: i\in \V \}$ the microscopic spin state of the system,  where $s_i\in \setS$ for all $i$. For Ising systems we would have $\setS=\{-1,1\}$.
A factor-tree is a special type of factor-graph in which there are no loops, see Figure \ref{figure:factor-tree}.  
The energy  $E_\nu(s)$ of each factor-node $\nu\in \F$ in a factor-graph is a specific function of the states of a subset $\partial_\nu\subseteq \V$ of the spins. 
We denote similarly  with  $\partial_i\subseteq \F$ the set of all factor-nodes connected to  variable-node $i$, i.e. all energy terms in (\ref{def:E})  that depend explicitly on spin $i$. Further examples of such systems are given in  Figure \ref{figure:squarelattice}.

From now on we will consider factor-trees only, as in Figure \ref{figure:factor-tree}.
We denote with with $\mu$  the factor-node at the root of the tree.
The full partition function $Z$ of the spin system (\ref{def:E}) on our factor tree can be written as
\begin{eqnarray}
Z&=&\sum_s\rme^{-\beta\sum_{\nu\in \F} E_\nu(s)}
=\sum_s\rme^{-\beta E_\mu(s)  -    \beta\sum_{\nu\in \F\setminus \{\mu\}} E_\nu(s)}
\nonumber
\\
&=&
\sum_{\{s_j, j\in\partial_\mu\}}\!\!\rme^{-\beta E_\mu(s)}\prod_{j\in \partial_\mu}
Z^{(0)}_{\mu j}[s_j].
\end{eqnarray}
Here $Z^{(0)}_{\mu j}[\tilde{s}_j]$ denotes the partition function of the sub-tree descending from node $j$, where $j$ descends from the root factor node $\mu$, 
\begin{eqnarray}
Z^{(0)}_{\mu j}[s_j]&=& \prod_{\nu\in\partial_j\setminus \mu}\Big(
\sum_{s_\ell,\ell\in \partial_\nu\setminus j}
\rme^{-    \beta E_\nu(s)}\prod_{\ell\in\partial_\nu\setminus j}Z^{(1)}_{\nu \ell}[s_\ell]\Big).
\end{eqnarray}
Continuation of this argument gives similar expressions  for any sub-tree function $Z_{\nu j}^{(r)}[\tilde{s}]$ at distance $r$ from the root in terms of the sub-tree-functions
at distance $r+1$, using the tree-like nature of the graph:
\begin{eqnarray}
Z^{(r)}_{\nu j}[s_j]&=&\prod_{\lambda\in\partial_j\setminus \nu}\Big(
\sum_{\{s_\ell, \ell\in\partial_\lambda \setminus j\}}\!\!\rme^{-\beta E_\lambda(s)}\prod_{\ell\in \partial_\lambda\setminus j}
Z^{(r+1)}_{\lambda \ell}[s_\ell]
\Big).
 \label{eq:Zr}
\end{eqnarray}
We can also calculate marginal spin probability distributions, starting from the top of the tree, and find
\begin{eqnarray}
P(\{s_j,j\in \partial_\mu \})&=&\frac{\rme^{-\beta E_\mu(s)}  \prod_{j\in \partial\mu}P^{(0)}_{\mu j}[s_j]}{\sum_{\{\tilde s_j, j\in\partial_\mu\}}\rme^{-\beta E_\mu(\tilde s)}  \prod_{j\in \partial\mu}P^{(0)}_{\mu j}[\tilde s_j]},\label{def:P}
\end{eqnarray}
where we have defined the probability distribution
\begin{eqnarray}
P^{(0)}_{\mu j}[s_j]&=&\frac{Z^{(0)}_{\mu j}[s_j]}{\sum_{\tilde s_j}Z^{(0)}_{\mu j}[\tilde s_j]}.
\label{def:P0}
\end{eqnarray}
$P^{(0)}_{\mu j}[s_j]$ is the marginal distribution of spin $j$ in the cavity graph, that is obtained upon removing the link from $j$ to factor node $\mu$. It follows directly from (\ref{eq:Zr}) that also this distribution can be computed recursively via
\begin{eqnarray}
\hspace*{-10mm}
P^{(r)}_{\nu j}[s_j]&=& \frac{\prod_{\lambda\in \partial j\setminus\nu}\Big(
 \sum_{\{s_\ell,\ell\in\partial_\lambda\setminus j\}}  
\rme^{-\beta  E_\lambda(s)}\prod_{\ell\in \partial\lambda\setminus j}
P^{(r+1)}_{\lambda \ell}[s_\ell]\Big) }
{\sum_{\tilde{s}_j}\prod_{\lambda\in \partial j\setminus\nu}\Big(
 \sum_{\{\tilde{s}_\ell,\ell\in\partial_\lambda\setminus j\}}  
\rme^{-\beta  E_\lambda(\tilde{s})}\prod_{\ell\in \partial\lambda\setminus j}
P^{(r+1)}_{\lambda \ell}[\tilde{s}_\ell]\Big) },
             \label{eq:Pr}
\end{eqnarray}
which is to be solved with boundary conditions $\{  P^{(\infty)}_{\nu j}[s_j]  \}$.  
From (\ref{def:P}) one obtains the ensemble averages of energies and magnetisations in layer $r=0$, i.e. for $i\in\partial_\mu$, via
 \begin{eqnarray}
 \langle E_\mu(s) \rangle&=&\sum_{\{ s_j,j\in\partial_\mu\}} P(\{s_j,j\in \partial_\mu \}) E_\mu(s),\label{eq:E}
\\
 \langle s_i \rangle&=&\sum_{\{ s_j,j\in\partial\mu\}} P(\{s_j, j\in \partial_\mu \}) s_i.      \label{eq:m}
\end{eqnarray}
If in a homogeneous system such as a regular ferromagnet  there exists an $r^\prime$ such that $P^{(r)}_{\mu j}[s]=P^c[s]$ for all $r\leq r^\prime$, i.e. if  the iteration    (\ref{eq:Pr}) for the spin distribution on the cavity graph has converged to an invariant measure $P[s]$,  
this defines a region in the tree where all factor-nodes and all variable-nodes have become equivalent. 
Any factor-node $\mu$ and any variable node $j$ belonging to this region will contribute 
\begin{eqnarray}
 \langle E_\mu(s) \rangle&=&
 \frac{\sum_{\{ s_j,j\in\partial_\mu\}} E_\mu(s) \rme^{-\beta E_\mu(s)}  \prod_{j\in \partial\mu}P^c[s_j]}{\sum_{\{s_j, j\in\partial_\mu\}}\rme^{-\beta E_\mu(s)}  \prod_{j\in \partial\mu}P^c[s_j]}
 \label{eq:E2}
\\
 \langle s_i \rangle&=&
 \frac{\sum_{\{ s_j,j\in\partial\mu\}} s_i  \rme^{-\beta E_\mu(s)}  \prod_{j\in \partial\mu}P^c[s_j]}{\sum_{\{ s_j, j\in\partial_\mu\}}\rme^{-\beta E_\mu(s)}  \prod_{j\in \partial\mu}P^c[s_j]}
 \label{eq:m2}
\end{eqnarray}
to the total energy and the total magnetization, respectively. In inhomogeneous systems this cannot happen;  the above averages would involve solutions of  (\ref{eq:Pr}) that would also depend on the realisation of the disorder, and one would have to turn to the population dynamics algorithm~\cite{Mezard2001} to  average out this disorder.

 For Ising spins, where  $s_i\in\{-1,1\}$ we can write (\ref{def:P0})
 in the following form, with parameters $h_{\mu j}(r)$ that in the spin-glass literature  are known as cavity fields \cite{Mezard2001}:
\begin{eqnarray}
 P^{(r)}_{\mu j}[s_j]&=&\frac{\rme^{\beta s_jh_{\mu j}(r)}}{2\cosh(\beta h_{\mu j}(r))}.
 \label{def:h}
\end{eqnarray}
 Using the  identity $\beta h_{\mu j}(r)=\frac{1}{2}\sum_{s_j}s_j\log P^{(r)}_{\mu j}[s_j]$ 
 in the left-hand side of (\ref{eq:Pr}) gives us a simple recursive equation for the cavity fields:
\begin{eqnarray}
\hspace*{-10mm}
\beta h_{\nu j}^{(r)}&=& \frac{1}{2 }\sum_{s_j}s_j\log \!\!
\prod_{\lambda\in\partial_j\setminus \nu}\!\!
\left(
\sum_{\{s_\ell,\ell\in\partial_\lambda\setminus j\}}
\!\!\rme^{-\beta E_\lambda(s)              +         \beta \sum_{\ell\in\partial_\lambda\setminus j}s_\ell h_{\lambda \ell}(r+1) }\right).
\label{eq:h}
\end{eqnarray}
The thermodynamics of homogeneous Ising systems of the type under consideration, where all spins are equivalent, would thus be governed by the solution of 
\begin{eqnarray}
\hspace*{-10mm}
\beta h&=& \frac{1}{2 }\sum_{s_j}s_j\log 
\prod_{\lambda\in\partial_j\setminus \nu}
\left(
\sum_{\{s_\ell,\ell\in\partial_\lambda\setminus j\}}\rme^{-\beta E_\lambda(s)  + \beta h\sum_{\ell\in\partial_\lambda\setminus j}  s_\ell }
\right)
\label{eq:h_homogeneous}
\end{eqnarray}
We can insert the solution of 
 (\ref{def:h}) into  equations  (\ref{eq:E2},\ref{eq:m2}), and use the fact that for homogeneous tree-like systems we could have chosen any factor node $\mu$ as our root, so that (\ref{eq:E2}) and (\ref{eq:m2}) must apply to all factor- and variable nodes. Hence we obtain for the energy density $E=\frac{1}{N}\sum_\mu \langle E_\mu(s)\rangle$ and the magnetisation per spin $m=\frac{1}{N}\sum_i \langle s_i\rangle$:
\begin{eqnarray}
E&=&\frac{1}{N}\sum_\mu
 \frac{\sum_{\{ s_j,j\in\partial_\mu\}} E_\mu(s) \rme^{-\beta E_\mu(s)+\beta h\sum_{j\in \partial\mu}
 s_j}
 }{\sum_{\{s_j, j\in\partial_\mu\}}\rme^{-\beta E_\mu(s)+\beta h\sum_{j\in \partial\mu} s_j}},
 \label{eq:E3}
\\
m&=&\frac{1}{N}\sum_i 
 \frac{\sum_{\{ s_j,j\in\partial_\mu\}} s_i  \rme^{-\beta E_\mu(s)+\beta h\sum_{j\in \partial\mu}s_j}}{\sum_{\{ s_j, j\in\partial_\mu\}}\rme^{-\beta E_\mu(s)+\beta h\sum_{j\in \partial\mu}s_j}}.
 \label{eq:m3}
\end{eqnarray}

As a simple example we can recover from the above equations the known results 
for the ferromagnetic Ising model on a Bethe lattice with connectivity $k$ (see e.g. \cite{Mezard2001}). Here $|\partial_\mu|=2$, $|\partial_j|=k$, and 
$E_\mu(s)=-Js_i s_j$ where $i,j\in\partial_\mu$.  The cavity field equation (\ref{eq:h_homogeneous}) now reduces to
 \begin{eqnarray}
h&=& \frac{1}{2\beta }\sum_{s}s\log \left[\sum_{s^\prime}\frac{\rme^{\beta J s s^\prime+\beta hs^\prime}}{2\cosh(\beta h)}\right]^{k-1}
\nonumber\\
&=& \frac{k\!-\!1}{\beta}\tanh^{-1}(\tanh(\beta h)\tanh(\beta J)).
\label{eq:Bethe-h}
\end{eqnarray}
 For the energy density (\ref{eq:E3}) and the magnetisation per spin  (\ref{eq:m3}) we find
\begin{eqnarray}
\hspace*{-10mm}
E&=&-\frac{1}{2}kJ
 \frac{\sum_{ss^\prime} ss^\prime\rme^{\beta Jss^\prime+\beta h(s+s^\prime)}}
{\sum_{ss^\prime} \rme^{\beta Jss^\prime+\beta h(s+s^\prime)}}
=
-\frac{1}{2}k J\frac{\tanh(\beta J )+\tanh(\beta h)^2}{1+\tanh(\beta J)\tanh(\beta h)^2},
\label{eq:Bethe-E}
\\
\hspace*{-10mm}
m&=& \frac{\sum_{ss^\prime} s\rme^{\beta Jss^\prime+\beta h(s+s^\prime)}}
{\sum_{ss^\prime} \rme^{\beta Jss^\prime+\beta h(s+s^\prime)}}=\frac{\tanh(\beta h)[1+\tanh(\beta J)]}{1+\tanh^2(\beta h)\tanh(\beta J)}.
\end{eqnarray}
Using (\ref{eq:Bethe-h}) the latter equation can be rewritten as
\begin{eqnarray}
m&=&\tanh(\beta hk/(k\!-\!1)).
\label{eq:Bethe-m}
\end{eqnarray}
For $k\in\{0,1, 2\}$ the system is always  paramagnetic ($m=0$). If $k\geq 3$ 
it is paramagnetic  for $\beta<\beta_c$ but ferromagnetic ($m\neq0$) for $\beta>\beta_c$, where 
$\beta_c=J^{-1}\tanh^{-1}(1/(k\!-\!1))$ is the critical inverse temperature of the system~\cite{Mezard2001}.

 \section{Ising models on hypercubic Bethe lattices \label{section:results}}
\begin{figure}[!t]
\setlength{\unitlength}{1mm}

\begin{center}
\begin{picture}(130,56)

\put(0,0){\includegraphics[width=120\unitlength]{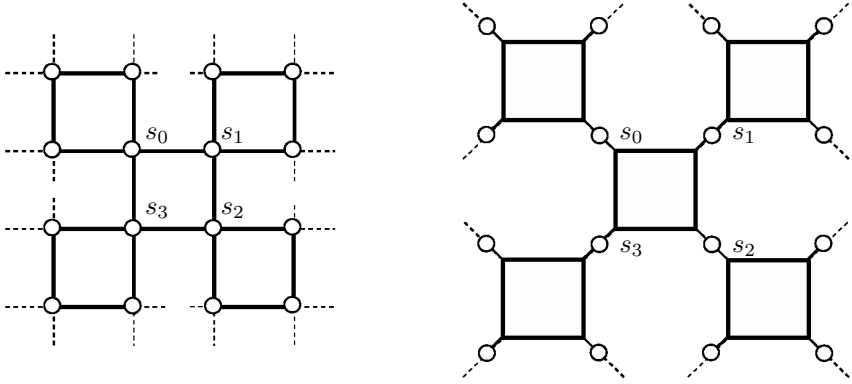}}

\put(23,35){$s_0$} \put(33,35){$s_1$}
\put(23,25){$s_3$} \put(33,25){$s_2$}

\put(86,35){$s_0$} \put(101,35){$s_1$}
\put(86,20){$s_3$} \put(101,20){$s_2$}

\end{picture}
\end{center}
\vspace*{-4mm}
\caption{Central part of the square lattice (on the left), which is also the central part of the hypercubic $d=2$  Bethe lattice, and its corresponding factor graph  representation (on the right).\label{figure:squarelattice} }
\end{figure}
We now turn to ferromagnetic Ising models on hypercubic Bethe lattices. These lattices are constructed recursively from a single $d$-dimensional hypercube, by attaching exactly  $2^d$ hypercubes to its corners, thereby producing the centre and the first shell of the lattice (see Figure \ref{figure:squarelattice}). The second shell is constructed by attaching  $2^d(2^d-1)$ hypercubes to the `available' corners in the first shell. This process of attaching  $d$-dimensional hypercubes to available corners is continued \emph{ad infinitum}. We note that each vertex in the hypercubic Bethe lattice  is shared between two adjacent hypercubes, ensuring that each vertex is connected to exactly $2d$ neighbouring vertices, exactly  as in conventional  $d$-dimensional lattices. To minimise notation clutter we will from now on choose units such that $J=1$, and transform $\beta h\to h$.

\subsection{Hypercubic Bethe lattice with $d=2$}

\begin{figure}[!t]
\setlength{\unitlength}{1mm}
\begin{center}
\begin{picture}(100,50)
\put(0,0){\includegraphics[height=45\unitlength]{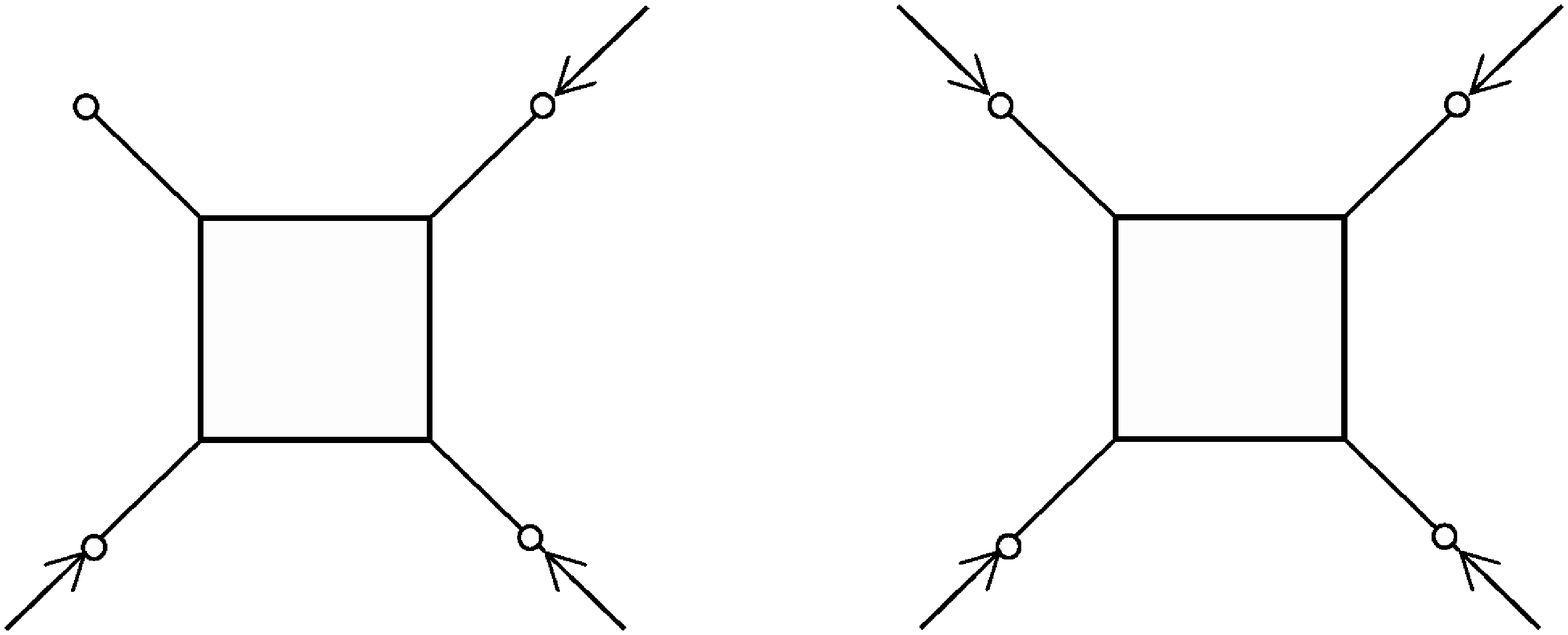}}
\put(10,35){$s_0$}  \put(38,35){$s_1$}
\put(10,8.5){$s_3$} \put(38,8.5){$s_2$}
\put(0,0){$h$}\put(42,0){$h$} \put(43,43){$h$}
\put(66,35){$s_0$}  \put(94,35){$s_1$}
\put(66,8.5){$s_3$} \put(94,8.5){$s_2$}
\put(55,0){$h$} \put(54,43){$h$}
\put(97,0){$h$} \put(98,43){$h$}
\end{picture}
\end{center}
\vspace*{-1mm}
\caption{Computations on the square Bethe lattice, i.e. the hypercubic Bethe lattice with $d=2$. Left: computation of the cavity fields. Right: evaluation of the marginal probability $P(s_0,\ldots, s_3)$, using the cavity fields. \label{figure:square} }
\end{figure}

\begin{figure}[t]
\setlength{\unitlength}{0.58mm}
\hspace*{1mm}
\begin{picture}(200,95)
\put(0,0){\includegraphics[height=90\unitlength,width=110\unitlength]{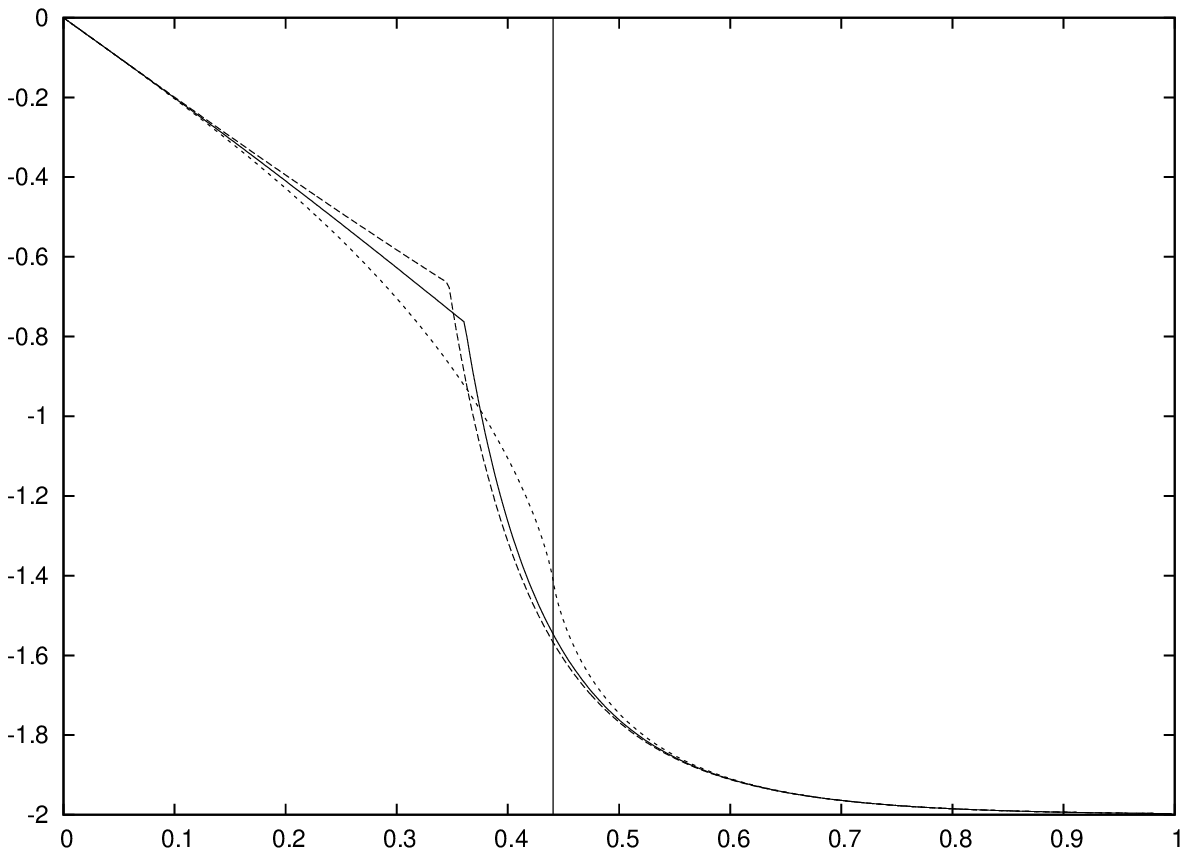}}

\put(57,37){\includegraphics[width=45\unitlength]{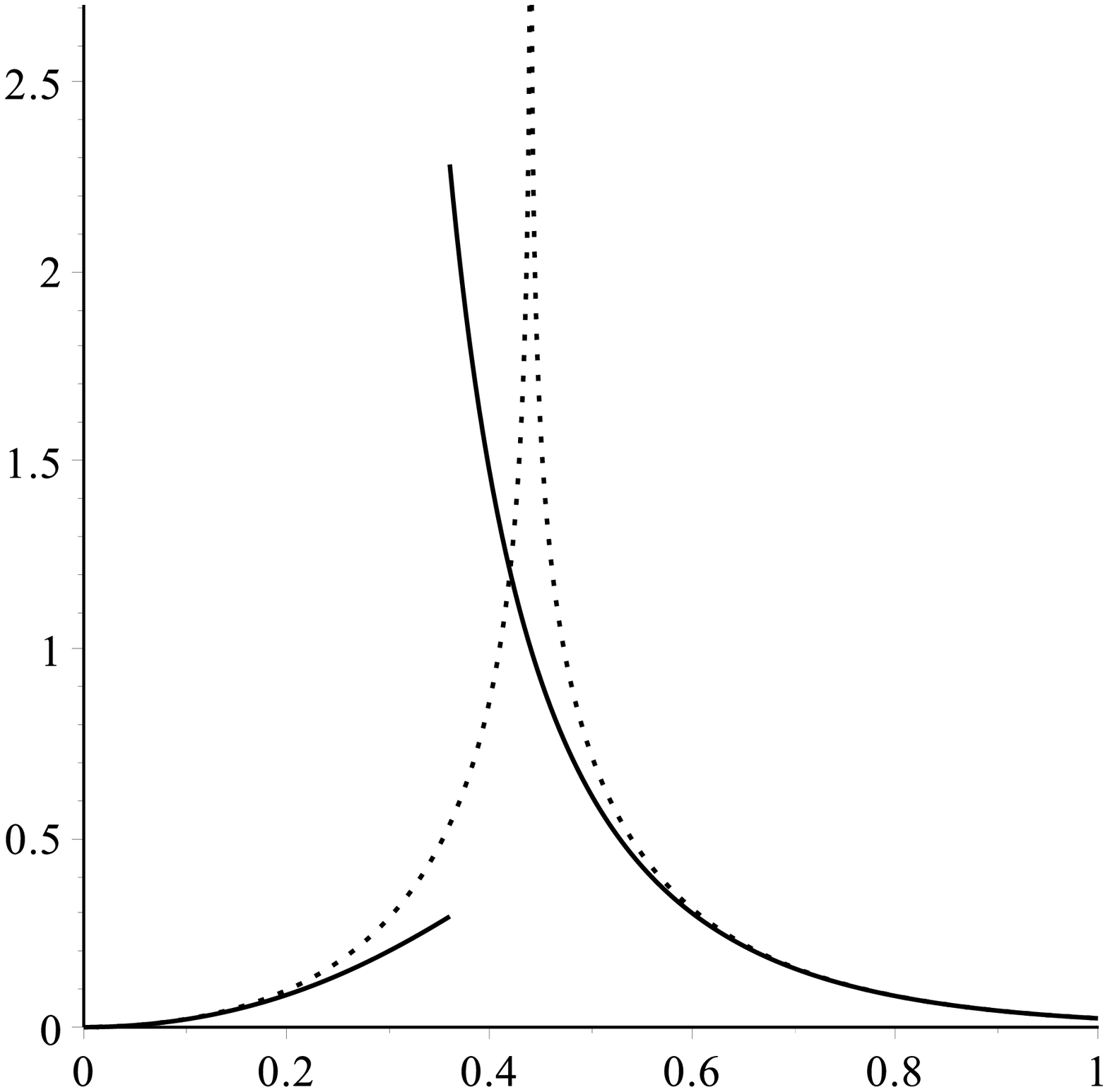}}
\put(53,71){\small{$C$}}
\put(115,0){\includegraphics[height=90\unitlength,width=110\unitlength]{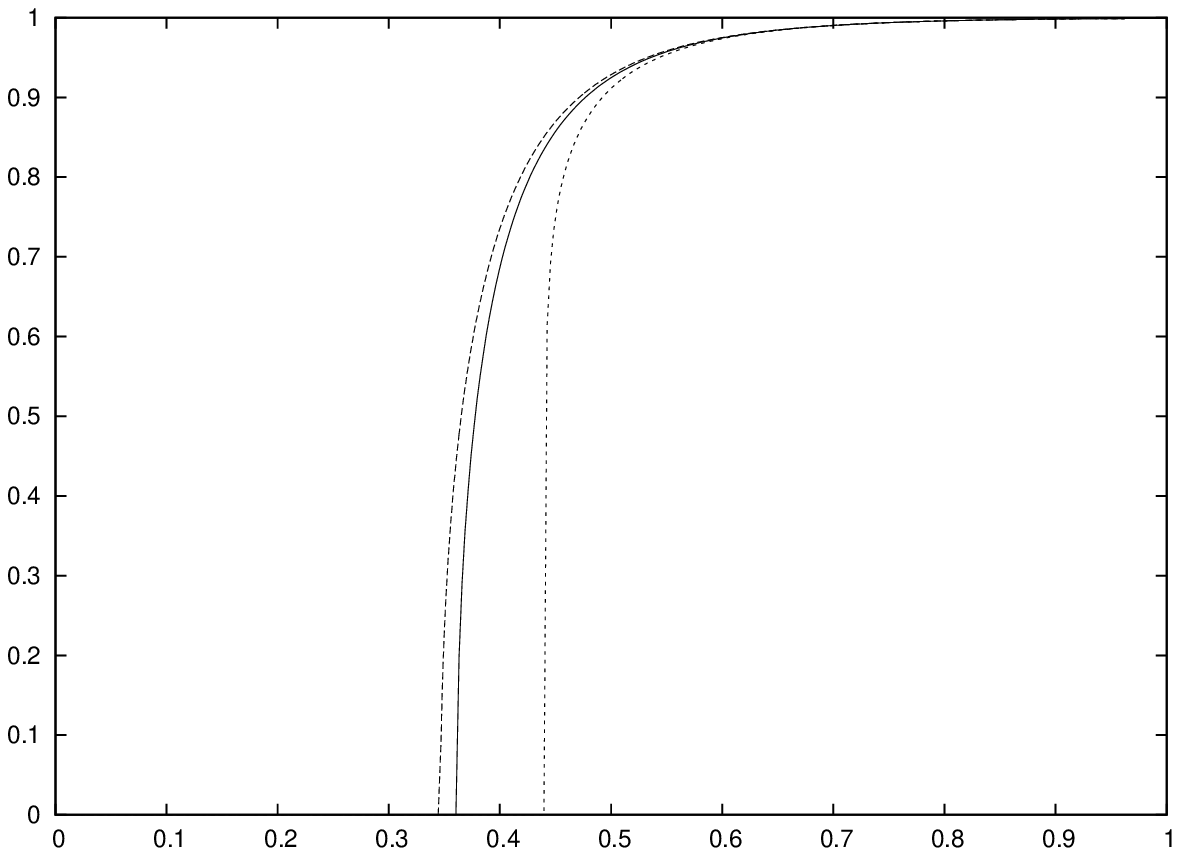}}
\put(55,-8){\small{$\beta$}}\put(170,-8){\small{$\beta$}}
\put(110,55){\small{$m$}}
\put(-5,55){\small{$E$}}
\end{picture}
\vspace*{6mm} 
\caption{Energy density $E$, specific heat $C$ and magnetisation per spin  $m$ as a function of the inverse temperature $\beta$ in $d=2$ lattices. Solid lines: square (hypercubic) Bethe lattice. Dashed lines: regular (tree-like) $k=4$ Bethe lattice. Dotted lines: the exact result for the Ising model in $d=2$ dimensions, with the critical inverse temperature $\beta_c=\frac{1}{2}\log(1+\sqrt{2})$ ~\cite{Onsager1944,Montroll1963} (vertical line in the left panel). } \label{figure:2d}
\end{figure}

Let us first study the simplest case of $d=2$. Here the hyper cubic Bethe lattice is constructed from squares, see Figure \ref{figure:squarelattice}.   
Each square contributes 
\begin{eqnarray}
E_{2d \square}(s)&=&- (s_{0}s_{1} + s_{1}s_{2} + s_{2}s_{3}+ s_{3}s_{0}) \label{eq:2d-E}
\end{eqnarray}
to the total energy of the system.  Using this in equation (\ref{eq:h}) gives us 
\begin{eqnarray}
 h(r)&=&\frac{1}{2} \log \left(  \frac{\sum_{\{s_j\}}\rme^{\beta\left[s_1 + s_1s_2 + s_2s_3+ s_3\right] +h(r+1)\sum_{j=1}^3s_j       }   }{\sum_{\{\tilde s_j\}}\rme^{\beta\left[-\tilde s_1 + \tilde s_1\tilde s_2 +\tilde s_2\tilde s_3 - \tilde s_3\right] +h(r+1)\sum_{j=1}^3\tilde s_j       }   }                     
 \right)\label{eq:2d-h}
\end{eqnarray}
for the cavity field acting on a spin $s_0$ living in the $r$-th shell of our lattice, see Figure \ref{figure:square}.  
If we solve this equation from a distant   boundary at $r\rightarrow\infty$,  we obtain the following equation for the cavity fields in the bulk of the system, describing the fixed-point of  the iterative map (\ref{eq:2d-h}):
\begin{eqnarray}
  h&=& \frac{1}{2} \log  \left( \frac {\rme^{-3h} +2\rme^{-h} +\rme^{-4\beta - h} +3\rme^{h} +\rme^{4\beta +3h} }
{\rme^{4\beta -3h} + 3\rme^{-h} +2\rme^{h} +\rme^{-4\beta +h} + \rme^{3h}     }\right).
\label{eq:2d-h-fp}
\end{eqnarray}
The cavity field $h$ acts on the spins living  in the central part of the square Bethe lattice,  and can be used  to compute the probability distribution 
of the four spins interacting on the square, see Figure \ref{figure:square}, being
\begin{eqnarray}
P(s_0,\ldots ,s_3)=\frac{                 \rme^{\beta\left[s_0s_1 + s_1s_2 + s_2s_3+ s_3s_0\right] +h\sum_{j=0}^3s_j }             }{   \sum_{\tilde{s}_0, \tilde{s}_1, \tilde{s}_2, \tilde{s}_3}\rme^{\beta\left[\tilde s_0\tilde s_1 + \tilde s_1\tilde s_2 + \tilde s_2\tilde s_3+ \tilde s_3\tilde s_0\right] +h\sum_{j=0}^3\tilde s_j                  }        }.
\label{eq:marginal}
\end{eqnarray}
Since our spin system is  homogeneous, this distribution can subsequently be used to compute the magnetisation per spin $m=\sum_{s_0,\ldots ,s_3}P(s_0,\ldots ,s_3)s_0$,  the energy density  $E=-2\sum_{s_0,\ldots ,s_3}P(s_0,\ldots ,s_3)s_0s_1$ and the specific heat $C=-\beta^2\frac{\partial}{\partial\beta}E$ (see  \ref{app:C} for details). The results are shown in Figure \ref{figure:2d}. 

The point $h=0$  is always a solution of equation (\ref{eq:2d-h-fp}), and corresponds to the $m=0$ paramagnetic (PM) state.  This solution becomes unstable at the critical inverse temperature $\beta_c=-\frac{1}{4}\log\left(\sqrt{ 5} -2\right)\approx 0.360909$, where a new solution $ h\neq0$ appears.  This new solution  is stable for   $\beta>\beta_c$, and   corresponds to the $m\neq0$ ferromagnetic (FM) state. The phase transition at $\beta_c$ is of second order  (see Figure \ref{figure:2d}). 

We compare these results with those of the ordinary $k=4$  Bethe lattice (\ref{eq:Bethe-m},\ref{eq:Bethe-E}), and with the exact results for  the $2$-dimensional ferromagnetic Ising  model ~\cite{Onsager1944,Montroll1963}. We then find that the transition point of our square Bethe lattice, i.e. $\beta_c\approx 0.360909$, is an improved estimate of the critical point  $\beta_c=\frac{1}{2}\log\left(1+\sqrt{2}\right)\approx 0.440687 $ \cite{Onsager1944} of the 2d Ising model, compared to  the estimate $\beta_c\approx0.346574 $ of the  ordinary Bethe lattice. We also observe this improvement over the ordinary Bethe estimates in terms of  the average  magnetisation and the energy density (see Figure \ref{figure:2d}), when comparing these quantities to the exact results for the 2d Ising model which are given by 
\begin{eqnarray}
\beta <\beta_c:&~~~&
m=0,\hspace*{14mm} E=   -2\sqrt{\frac{1\!+\!\kappa}{\kappa}} \Big(   \frac{\kappa-1}{\pi} \mathcal{K}(\kappa) +\frac{1}{2} \Big)
\label{eq:2d-lattice-mE-high-T}    
\\
\beta >\beta_c:&~~~&
m=(1\!-\!\kappa^2)^{\frac{1}{8}},\hspace*{3mm} E=   -2\sqrt{1\!+\!\kappa}    \Big(   \frac{1-\kappa}{\pi} \mathcal{K}(\kappa) +\frac{1}{2}               \Big)
\label{eq:2d-lattice-mE-low-T}
\end{eqnarray}
Here $\kappa=\sinh^{-2}(2\beta)$, and  $\mathcal{K}(\kappa)$ is the complete elliptic integral of the first kind~\cite{Montroll1963}. Finally, we note that the specific heat $C$ of the square Bethe lattice has a jump at $\beta_c=-\frac{1}{4}\log\left(\sqrt{ 5} -2\right)$ (see Figure \ref{figure:2d} and \ref{app:C})  this is in contrast to the conventional square lattice where the specific heat is diverging at $\beta_c=\frac{1}{2}\log\left(1+\sqrt{2}\right) $. This behaviour of the specific heat is also found in $d=3,4$ Bethe lattices  (see  \ref{app:C} for details) studied in the next two sections.

\subsection{Hypercubic Bethe lattice with $d=3$}

\begin{figure}[t]
\setlength{\unitlength}{0.58mm}
\hspace*{1mm}
\begin{picture}(200,95)
\put(0,0){\includegraphics[height=90\unitlength,width=110\unitlength]{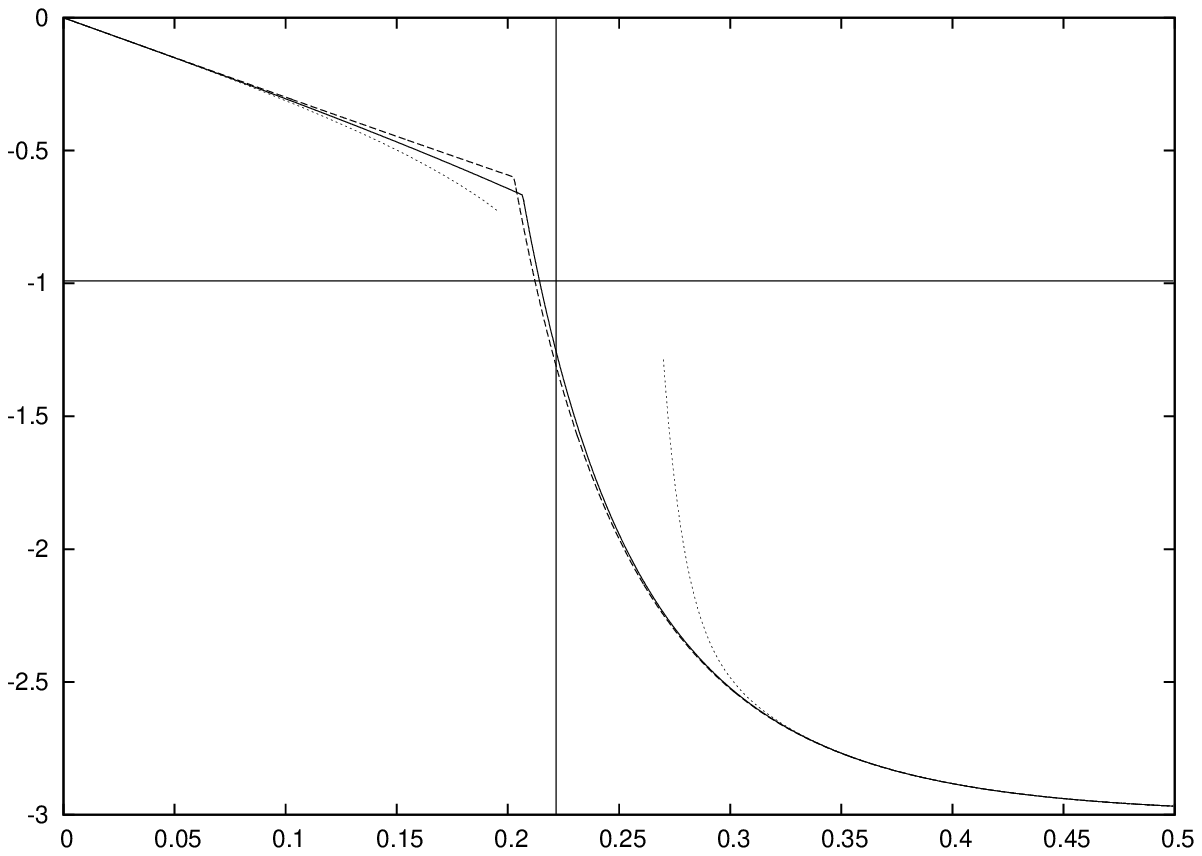}}


\put(115,0){\includegraphics[height=90\unitlength,width=110\unitlength]{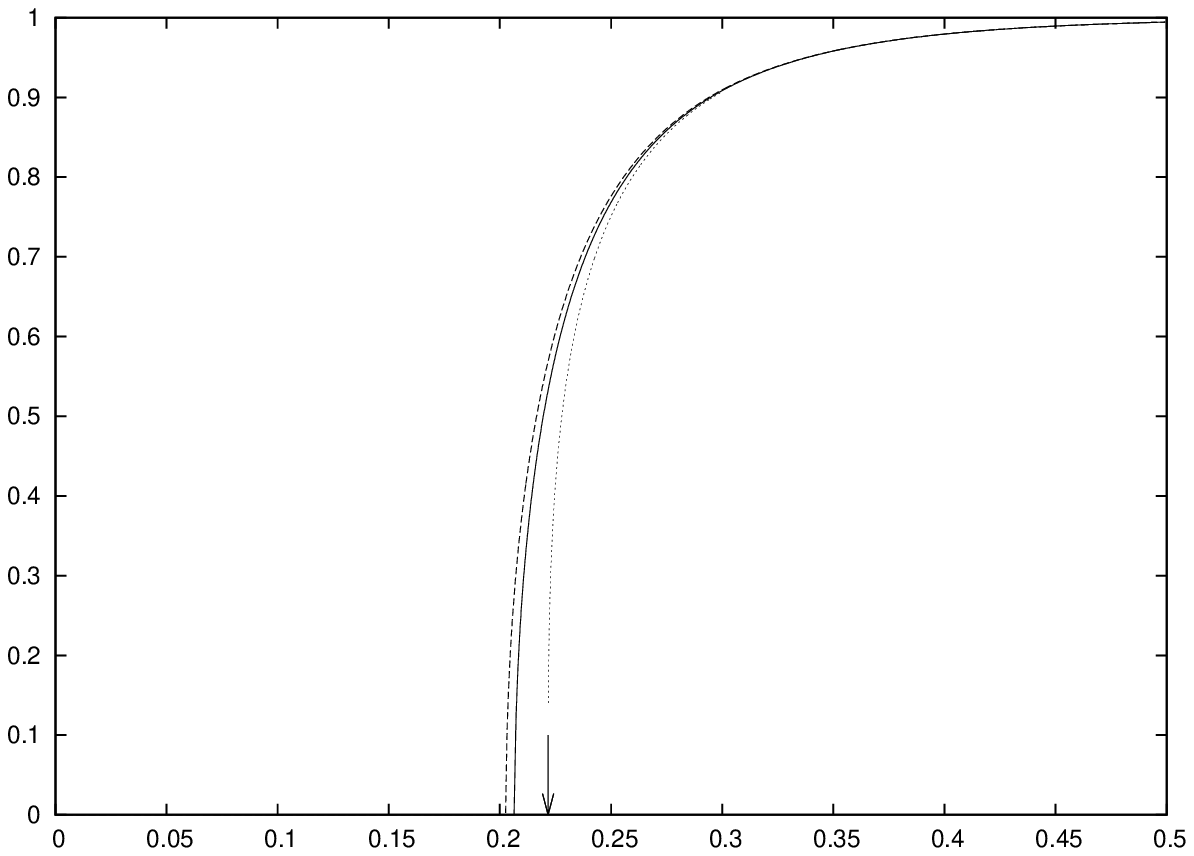}}
\put(55,-8){\small{$\beta$}}\put(170,-8){\small{$\beta$}}
\put(110,55){\small{$m$}}
\put(-5,55){\small{$E$}}
\end{picture}
\vspace*{6mm} 
\caption{
Energy density $E$ and magnetisation per spin  $m$ as a function of the inverse temperature $\beta$ for $d=3$ lattices. Solid lines: cubic  Bethe lattice. Dashed lines: regular (tree-like) $k=6$ Bethe lattice. 
The results for the true Ising model in $d=3$ dimensions are low temperature data for  $m$ (dotted line) obtained by MC simulations in~\cite{Talapov1996}, with the critical inverse temperature estimate $\beta_c=0.2216544 \pm 3\times 10^{-7}$~\cite{Talapov1996} (arrow and vertical line), predictions for $E$ obtained from high temperature~\cite{Arisue2003} and low temperature \cite{Bhanot1994} series (both as dotted lines). The  energy density $E=-0.99063\pm3\times10^{-5}$ (horizontal  line), assuming that  $\beta_c=0.2216546 $, was computed by MC simulations in~\cite{Haggkvist2007}. 
 } \label{figure:3d}
\end{figure}

In the Ising model on the hypercubic $d=3$ Bethe lattice,  a cube, which one can regard as an `upper' square (formed of variable nodes 0,1,2,3) that is is connected to a `lower' square (formed of variable nodes 4,5,6,7), contributes
\begin{eqnarray}
E_{3d\,\square}(s)&=&-(s_0   s_1 +  s_1 s_2 +  s_2   s_3 +  s_0   s_3) \nonumber\\
&&-  (s_4   s_5 +  s_5   s_6 +  s_6 s_7 +  s_7   s_4 )\nonumber\\
&&- ( s_0 s_4  + s_1s_5 +  s_2s_6  +   s_3s_7 ) \label{eq:3d-E1}
\end{eqnarray}
to the total energy. Substituting this into (\ref{eq:h}) gives us after some straightforward algebra the following equation for the cavity field $h$:
\begin{eqnarray}
h&=&\frac{1}{2}\ln \big[\big(
5 {\rme^{-6 \beta+h}}+12 {\rme^{3 h}}+7 {
\rme^{6 \beta+5 h}}+{\rme^{-12 \beta-h}}+{
\rme^{6 \beta-7 h}}
\nonumber
\\
&&+3 {\rme^{4 \beta-5 h}
}+4 {\rme^{-5 h}}+9 {\rme^{2 \beta-3 h}}+
3 {\rme^{4 \beta-h}}+9 {\rme^{-2 \beta-3 {\it h
}}}\nonumber\\
&&+16 {\rme^{-h}}+15 {\rme^{2 \beta+h}}+3
 {\rme^{-6 \beta-3 h}}+15 {\rme^{-4 \beta-{\it 
h}}}+15 {\rme^{-2 \beta+h}}\nonumber\\
&&+9 {\rme^{4 \beta+3 
h}}+{\rme^{12 \beta+7 h}}
\big)\big(     {\rme^{12 \beta-7 h}}+7 {\rme^{6 \beta-5 {\it h
}}}\nonumber\\
&&~~+9 {\rme^{4 \beta-3 h}}+12 {\rme^{-3 h
}}+15 {\rme^{2 \beta-h}}+15 {\rme^{-2 \beta-{\it 
h}}}+5 {\rme^{-6 \beta-h}}\nonumber\\
&&~~+15 {\rme^{-4 \beta+{
\it h}}}+16 {\rme^{h}}+9 {\rme^{-2 \beta+3 {\it 
h}}}+{\rme^{-12 \beta+h}}+3 {\rme^{-6 \beta+3 {
\it h}}}\nonumber\\
&&~~+3 {\rme^{4 \beta+h}}+9 {\rme^{2 \beta+3
 h}}+4 {\rme^{5 h}}+3 {\rme^{4 \beta+5 h}}+{\rme^{6 \beta+7 h}}\big)^{-1}\big].
\label{eq:3d-cavity}
\end{eqnarray}
The paramagnetic solution  $h=0$ of this equation becomes unstable at $\beta_c\approx0.206633$, and for $\beta > \beta_c$ equation (\ref{eq:3d-cavity}) has two $h\neq0$ ferromagnetic solutions. 
Substitution of the the factor node energy function (\ref{eq:3d-E1}) into equations (\ref{eq:m}) and (\ref{eq:E}) gives us the corresponding magnetisation per spin and the energy density: 
\begin{eqnarray}
m&=&\big(
6 {\rme^{6 \beta+6 h}}-{\rme^{12 \beta-8 {\it h
}}}-6 {\rme^{6 \beta-6 h}}-6 {\rme^{4 \beta-4 {
\it h}}}-8 {\rme^{-4 h}}
\nonumber\\
&&-6 {\rme^{2 \beta-2 {
\it h}}}-6 {\rme^{-2 \beta-2 h}}-2 {\rme^{-6 
\beta-2 h}}+6 {\rme^{2 \beta+2 h}}
\label{eq:3d-m}
\\
&&+6 {
\rme^{-2 \beta+2 h}}+6 {\rme^{4 \beta+4 h
}}+2 {\rme^{-6 \beta+2 h}}+8 {\rme^{4 h}}
+{\rme^{12 \beta+8 h}}\big)/\mathcal{N},
\nonumber
\\[1mm]
E&=&-3\big(
4 {\rme^{6 \beta+6 h}}+{\rme^{12 \beta-8 h}}+4 {\rme^{6 \beta-6 h}}+4 {\rme^{4 \beta-4 {
\it h}}}+4 {\rme^{2 \beta-2 h}}
\nonumber\\
&&~~+2 {\rme^{4 
\beta}}-4 {\rme^{-2 \beta-2 h}}-4 {\rme^{-6 \beta-
2 h}}-10 {\rme^{-4 \beta}}+4 {\rme^{2 \beta+2 h}}
\label{eq:3d-E2}
\\
&&~~-4 {\rme^{-2 \beta+2 h}}+4 {\rme^{4 
\beta+4 h}}-2 {\rme^{-12 \beta}}-4 {\rme^{-6 \beta
+2 h}}+{\rme^{12 \beta+8 h}}
\big)/\mathcal{N},
\nonumber
\end{eqnarray}
where
\begin{eqnarray}
\mathcal{N}&=& 32+8 {\rme^{6 \beta+6 h}}+{\rme^{12 \beta-8 h}}+8 {\rme^{6 \beta-6 h}}+12 {\rme^{4 \beta-4
 h}}+16 {\rme^{-4 h}}
 \nonumber
 \\
&&~~+24 {\rme^{2 \beta-2
 h}}+6 {\rme^{4 \beta}}+24 {\rme^{-2 \beta-2 h}}+8 {\rme^{-6 \beta-2 h}}+30 {\rme^{-4 
\beta}}\nonumber\\
&&~~+24 {\rme^{2 \beta+2 h}}+24 {\rme^{-2 \beta
+2 h}}+12 {\rme^{4 \beta+4 h}}+2 {\rme^{-
12 \beta}}+8 {\rme^{-6 \beta+2 h}}\nonumber\\
&&~~+16 {\rme^{4 h}}+{\rme^{12 \beta+8 h}}.
\label{eq:3d-Norm}
\end{eqnarray}
In Figure \ref{figure:3d} we compare the above values for $m$ and $E$ with those of the simple tree-like $k=6$ Bethe lattices and with the predictions (obtained via simulations and expansions) for the true $d=3$ Ising model. Again we find that the hypercubic Bethe lattice is a more accurate proxy for the true Ising system than the tree-like Bethe lattice with the same coordination number. 

\subsection{Hypercubic Bethe lattice with $d=4$}
%

The calculation for $d=4$ is similar to $d=3$ but more tedious. In the Ising model on the hypercubic $d=4$ Bethe lattice the tesseract (the $d=4$ equivalent of a cube in $d=3$) can be viewed as a $3d$ cube (A) connected to another $3d$ cube (B) by 8 edges. It thus contributes to the total energy an amount 
\begin{eqnarray}
E_{4d\,\square}(s)&=& E_A(s)+E_B(s)+E_{AB}(s)
\label{eq:4d-E1}
\end{eqnarray}
where
\begin{eqnarray}
E_A(s)&=&-\big(s^A_0   s^A_1 +  s^A_1 s^A_2 +  s^A_2   s^A_3 +  s^A_0   s^A_3
+s^A_4   s^A_5 +  s^A_5   s^A_6 \nonumber
\\
&&  ~~~~+  s^A_6 s^A_7 +  s^A_7   s^A_4+
 s^A_0 s^A_4  + s^A_1s^A_5 +  s^A_2s^A_6  +   s^A_3s^A_7 \big),
 \\[1mm]
E_B(s)&=&-\big(s^B_0   s^B_1 +  s^B_1 s^B_2 +  s^B_2   s^B_3 +  s^B_0   s^B_3
+s^B_4   s^B_5 +  s^B_5   s^B_6 \nonumber
\\
&&  ~~~~+  s^B_6 s^B_7 +  s^B_7   s^B_4+
 s^B_0 s^B_4  + s^B_1s^B_5 +  s^B_2s^B_6  +   s^B_3s^B_7 \big),
\\
E_{AB}(s)
&=&-\sum_{j=0}^7s^A_js^B_j .
\end{eqnarray}
The resulting full equations for the cavity field $h$, the energy density $E$ and the magnetisation per spin $m$ are given in \ref{app:formulas}. 
Bifurcation analysis around the trivial solution $h=0$ of the equation for the cavity field now reveals a second order transition 
from a paramagnetic state ($m=0$, for $\beta<\beta_c$) to  a ferromagnetic state ($m\neq 0$, for $\beta>\beta_c$)  at $\beta_c\approx 0.145361$.

\begin{figure}[t]
\setlength{\unitlength}{0.58mm}
\hspace*{1mm}
\begin{picture}(200,95)
\put(0,0){\includegraphics[height=90\unitlength,width=110\unitlength]{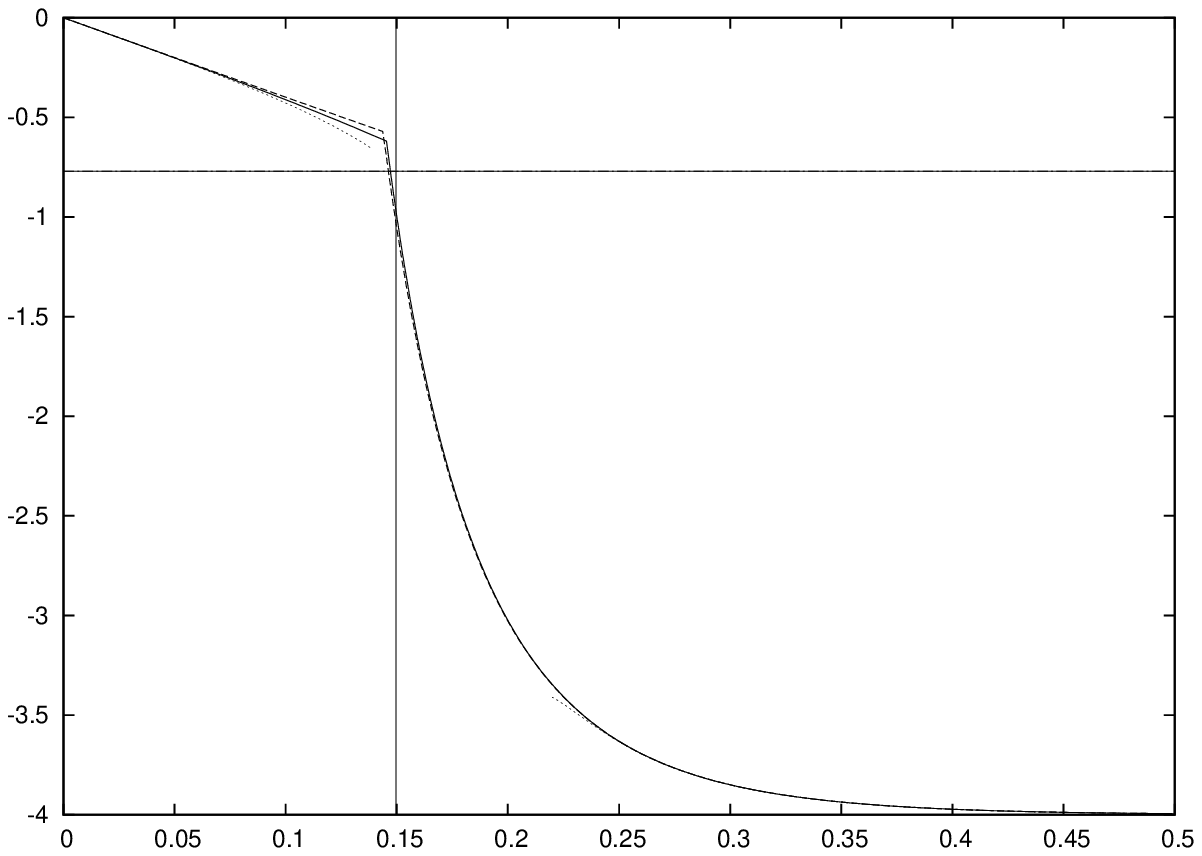}}


\put(115,0){\includegraphics[height=90\unitlength,width=110\unitlength]{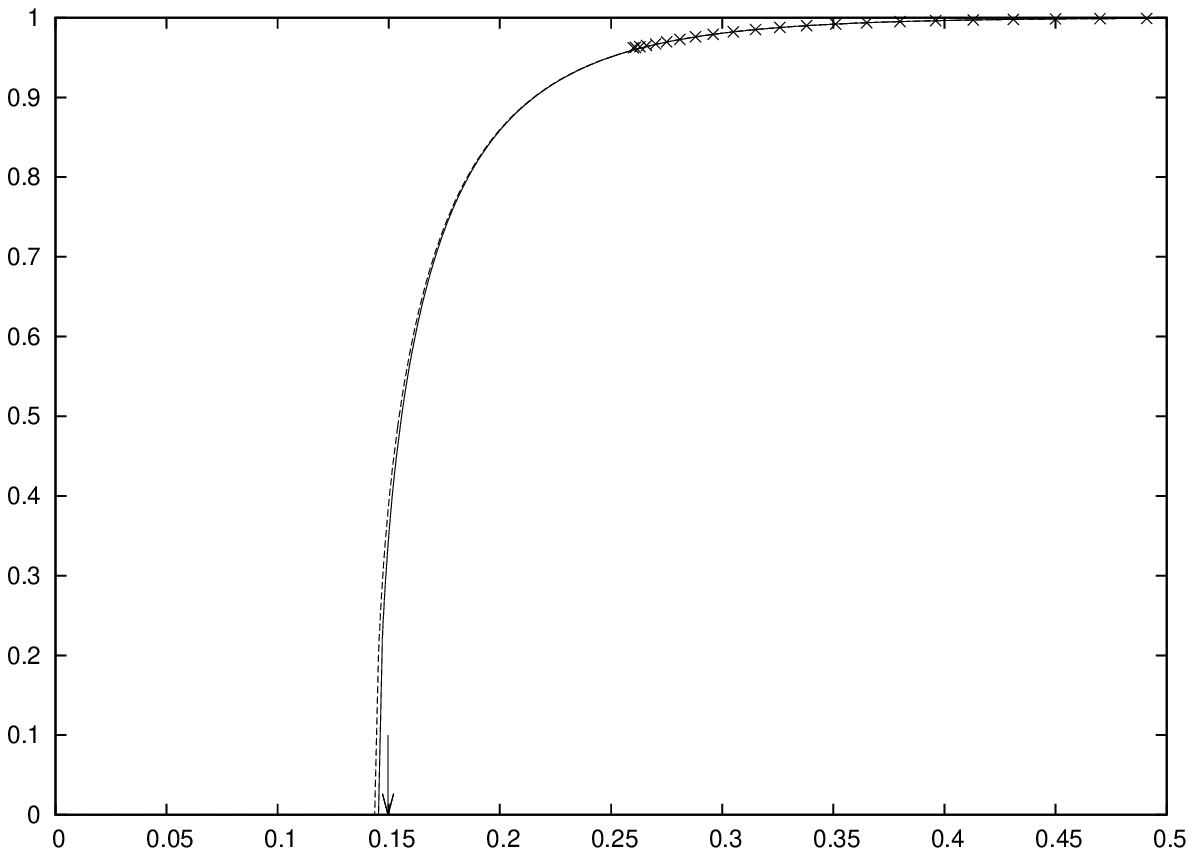}}
\put(55,-8){\small{$\beta$}}\put(170,-8){\small{$\beta$}}
\put(110,55){\small{$m$}}
\put(-5,55){\small{$E$}}
\end{picture}
\vspace*{6mm} 
\caption{
Energy density $E$ and magnetisation per spin  $m$ as a function of the inverse temperature $\beta$ for $d=4$ lattices. Solid lines: hypercubic (tesseract) Bethe lattice. Dashed lines: regular (tree-like) $k=8$ Bethe lattice. 
The results shown for the Ising model in $d=4$ dimensions are obtained from low temperature series for $E$ \cite{Bhanot1994}  (dotted line), low temperature series for $m$ \cite{Vohwinkel1993}  ($\times$), and high temperature series for $E$ \cite{Sykes1979} (dotted line). The  energy density at the critical point $E=-0.77053\pm4\times10^{-5}$ (horizontal  line) and the critical inverse temperature $\beta_c=0.1496947 \pm 5\times 10^{-7}$ were computed in MC simulations \cite{Lundow2009} (arrow and vertical line). 
 } \label{figure:4d}
\end{figure}

In Figure \ref{figure:4d} we compare the values found for $m$ and $E$ with those of the simple tree-like $k=8$ Bethe lattices and with the predictions (obtained via simulations and expansions) for the true $d=4$ Ising model. As was the case with $d=2,3$ we find that, although the differences between the different model versions become smaller as $d$ increases,  also in $d=4$ the hypercubic Bethe lattice is a more accurate proxy for the true Ising lattice than the tree-like Bethe lattice with the same coordination number.

\section{Discussion\label{section:discussion}}

In this paper  we introduced hypercubic Bethe lattices, which are constructed from the cells of regular $d$-dimensional cubic lattices, and we analysed the equilibrium properties of spin systems defined on such lattices. These topologies can be seen as a further generalisations of ordinary Bethe lattices,  that, unlike tree-like graphs, retain many of the loops of the interaction topologies of more realistic spin systems.  

We used the Bethe-Peierls method to derive equations for the magnetisation per spin and the energy density for Ising spin systems on hypercubic Bethe lattices, using the factor-tree representation. With these equations we computed phase diagrams for the ferromagnetic Ising model  on the hypercubic Bethe lattice with $d=2,3$ and $4$. The results for the critical temperatures are summarised in Table 
\ref{table:Tc}, and compared with the values found for  true cubic lattices and for ordinary Bethe lattices. 
Hypercubic Bethe lattices are found to be more accurate proxies for the true $d$-dimensional lattices than regular (tree-like)  $k=2d$ Bethe lattices, in terms of the predicted transition temperatures  and the values of observables. However, for $d\geq 4$ one finds, as expected, that the differences between the predictions of all three models become increasingly small.  We expect that for $N\to\infty$ the hypercubic Bethe lattice is  equivalent to its random graph version, which is a maximally random graph constructed from hypercubes, and with vertices of equal degree,  at least for ferromagnets~\cite{Johnston1998}.

\begin{table}[t]
\vspace*{3mm}
\hspace*{25mm}
\begin{tabular}{l|ccc}
\hline
& $T_c ~(d\!=\!2)$ & $T_c ~(d\!=\!3)$ & $T_c ~(d\!=\!4)$ 
\\
\hline\hline
{\rm true cubic lattice}            &   2.2692  & 4.5115  & 6.6802  \\
{\rm hypercubic Bethe lattice} &   2.7708  & 4.8395  & 6.8794 \\
{\rm $k=2d$ Bethe lattice}      &   2.8854  & 4.9326  & 6.9521
\\
\hline
\end{tabular}
\caption{Critical temperatures $T_c=\beta_c^{-1}$ of Ising models on true cubic lattices, hypercubic Bethe lattices and $k=2d$ Bethe lattices, for $d=2,3,4$. The values for the true $d=2$ lsing lattice and for all hypercubic and ordinary  Bethe lattices are calculated analytically. The values of $T_c$ for the cubic lattice with $d=3,4$ are computed in MC simulations (see captions of previous figures for references). }
\label{table:Tc}
\end{table}

Our results for the Ising model were obtained by computing  the sums in the equations (\ref{eq:h},\ref{eq:E2},\ref{eq:m2}) directly.  However, even for the hypercube in  $d=5$ (the penteract) this involves $O(2^{32})$ summations (since the number of corners in the $d$-dimensional  hypercube is $2^d$), which although in principle easy becomes painful in practice. 
 
 One interesting  future direction following this work would be to consider the one loop correction method of \cite{Montanari2005}, which was introduced and used to improve the Bethe ($k=2d$) estimates for the critical temperature of the ferromagnetic Ising model on the $d$-dimensional  lattice. Generalisation of their method to the hypercubic Bethe lattice may further improve our estimates of $T_c$.  In addition it would be interesting  to study a spin-glass on the hypercubic random regular graph;  the spin-glass model on loopy graphs, to the best of our knowledge, was studied only for the Husimi lattices~\cite{Lage2008, Yokota2008}  and for a single $d$-dimensional hypercube~\cite{Fernandez2010} in the limit of large $d$.  The non-negligible frustration in spin-glasses on random graphs is due to the presence of long $O(N)$ loops~\cite{Mezard2001}, but in our present  model it would already enter via the short loops that are present. 

\section*{Acknowledgements}

The authors would like to thank  Alessia Annibale for interesting discussions.

\section*{References}


\appendix
\section{Ising models on hypercubic Bethe lattices: Specific heat\label{app:C}}
Using the energy (\ref{def:E}) in the definition $C(T)= \frac{\partial}{\partial T}\langle E(s)/N\rangle$  gives us the  specific  heat (density) equation 

\begin{eqnarray}
C(T)
&=& \frac{1}{N} \sum_{\nu\in \F}  \frac{1}{ T^2}\left[\langle E_\nu^2(s) \rangle -\langle E_\nu(s) \rangle^2\right]\label{def:C(T)}
\end{eqnarray}
or equivalently 
\begin{eqnarray}
C(\beta)&=& \frac{1}{N} \sum_{\nu\in \F}   -\beta^2\frac{\partial}{\partial \beta}\langle E_\nu(s) \rangle \label{eq:C(1/T)}.
\end{eqnarray}
The Ising model on a hypercubic Bethe lattice  is a homogeneous system, i.e. all its nodes are equivalent, so  $\langle E(s)/N\rangle =\frac{M}{N} \langle E_0(s) \rangle$, where $M=N2^{1-d}$ and 
\begin{eqnarray}
\langle E_0(s)\rangle&=&
 \frac{\sum_{\{ s_j\}} E_0(s)\, \rme^{-\beta E_0(s) + h\sum_{j}
 s_j}
 }{\sum_{\{s_j\}}\rme^{-\beta E_0(s) + h\sum_{j} s_j}}. \label{eq:<E0>}
\end{eqnarray}
The magnetization $m=\frac{1}{N}\sum_{i=1}^N\langle s_i\rangle=\langle s_0\rangle$ is given by the equation
\begin{eqnarray}
m&=&
 \frac{\sum_{\{ s_j\}} s_0 \rme^{-\beta E_0(s) + h\sum_{j}
 s_j}
 }{\sum_{\{s_j\}}\rme^{-\beta E_0(s) + h\sum_{j} s_j}} \label{eq:<s0>}
\end{eqnarray}
and the cavity field $h$ satisfies the equation  
\begin{eqnarray}
h&=&f_\beta(h)= \frac{1}{2 }\sum_{s_0}s_0\log 
\left(
\sum_{\{s_j\}}\rme^{-\beta E_0(s)  + h\sum_{j\neq0}  s_j}
\right)\label{def:f(h)}.
\end{eqnarray}
The magnetization and the cavity field are related by the equation 
\begin{eqnarray}
m&=&\tanh(2h)\label{eq:m(h)}
\end{eqnarray}
which follows from the equations (\ref{eq:<s0>}), (\ref{def:f(h)}) and the equality $\tanh^{-1}(x)=\frac{1}{2}\log\left(\frac{1+x}{1-x}\right)$ valid for $x\in(-1,1)$. Using the equations (\ref{eq:<E0>}) and (\ref{def:f(h)}) in the formula (\ref{eq:C(1/T)}) gives us the specific heat 
\begin{eqnarray}
C(\beta)&=& -2^{1-d}  \beta^2\frac{\partial}{\partial \beta}\langle E_0(s) \rangle\label{eq:C-hyper}\\
&=&2^{1-d}\beta^2  \Big\{\left[\langle E_0^2(s) \rangle -\langle E_0(s) \rangle^2\right] \nonumber\\
&&-  \sum_{j}\left[\langle E_0(s) s_j\rangle -\langle E_0(s) \rangle \langle s_j \rangle\right]\frac{\partial h}{\partial\beta}\Big\}.\nonumber
\end{eqnarray}

Let us now assume that $E_0(s)=E_0(-s)$ (this property is satisfied in our model) and consider 
\begin{eqnarray}
f_\beta(-h)&=&\frac{1}{2 }\sum_{s_0}s_0\log \left(\sum_{\{s_j\}}\rme^{-\beta E_0(s)   - h\sum_{j\neq0}  s_j }\right)\label{eq:f(0)}\\
&=& \frac{1}{2 }\sum_{s_0}s_0\log 
\left(
\sum_{\{s_j\}}\rme^{-\beta E_0(-s) - h\sum_{j\neq0}  s_j}
\right)\nonumber\\
&=& -\frac{1}{2 }\sum_{s_0}s_0\log 
\left(
\sum_{\{s_j\}}\rme^{-\beta E_0(s) + h\sum_{j\neq0}  s_j}
\right)=-f_\beta(h)\nonumber.
\end{eqnarray}
Thus $f_\beta(h)$ is an odd function of $h$ and hence $h=0$ is a solution of the cavity field equation (\ref{def:f(h)}). Furthermore, let us assume that $E_0(s)$ is ferromagnetic, i.e. $E_0(s)=-\sum_A J_A s_A$ ($A$ is a set of indices and $s_A=\prod_{i\in A}s_i$ with $J_A\geq0$), and $h\geq0$ then $f_\beta(h)$ is a monotonic non-decreasing function of $h$. To show this we first note that 
\begin{eqnarray}
f_\beta(h)&=&\frac{1}{2}\log\left(\frac{P^c[+1]}{P^c[-1]}\right)\label{eq:h(P)}\\
P^c[s_0]&=&\frac{\sum_{\{s_j\}}\rme^{-\beta E_0(s)  + h\sum_{j\neq0}  s_j}}{ \sum_{\tilde s_0}\sum_{\{\tilde s_j\}}\rme^{-\beta E_0(\tilde s)  + h\sum_{j\neq0}  \tilde s_j}}\label{eq:P^c}
\end{eqnarray}
and $P^c[s_0]=\frac{1}{2}[1+s_0\langle s_0\rangle_c]$, where the average $\langle \cdots\rangle_c$ is generated by the Boltzmann weight  $\rme^{-\beta E_0(s)  + h\sum_{j\neq0}  s_j}$,  then 
\begin{eqnarray}
\frac{\partial}{\partial h}f_\beta(h)&=&\frac{\partial}{\partial h} \frac{1}{2}\log\left(\frac{1+\langle s_0\rangle_c     }{1-\langle s_0\rangle_c}\right) = \frac{1}{1-\langle s_0\rangle_c^2}\frac{\partial}{\partial h}\langle s_0\rangle_c\label{eq:dfdh}\\
%
%
&=& \frac{1}{1-\langle s_0\rangle_c^2}\sum_{j\neq0}\left[\langle s_0s_j\rangle_c - \langle s_0\rangle_c \langle s_j\rangle_c\right].\nonumber
\end{eqnarray}
Secondly the inequality  $\langle s_0s_j\rangle_c - \langle s_0\rangle_c \langle s_j\rangle_c\geq0 $ is true by the Griffiths-Kelly-Sherman (GKS) theorem~\cite{Domb1972} and hence $\frac{\partial}{\partial h}f_\beta(h)\geq0$. For $h=0$ the cavity  magnetizations $\langle s_0\rangle_c=\tanh(h)$ vanishes and  $\frac{\partial}{\partial h}f_\beta(h)=\sum_{j\neq0} \langle s_0s_j\rangle_c$. Furthermore, the gradient of $f_\beta(h)$ at $h=0$ is a monotonic increasing function of $\beta$. This follows from the calculation
\begin{eqnarray}
\frac{\partial}{\partial \beta}\frac{\partial}{\partial h}f_\beta(h)&=&\sum_{j\neq0} \frac{\partial}{\partial \beta}\langle s_0s_j\rangle_c\label{eq:dfdb}\\
&=&\sum_{j\neq0}\left[ -\langle s_0s_jE_0(s) \rangle_c + \langle s_0s_j \rangle_c \langle E_0(s) \rangle_c\right]\nonumber\\
&=&\sum_{j\neq0}\sum_A J_A\left[ \langle s_0s_js_A \rangle_c - \langle s_0s_j \rangle_c \langle s_A \rangle_c\right]\geq0,\nonumber
\end{eqnarray}
where the inequality is true by the  GKS theorem. For a $d$-dimensional hypercube we have that $0\leq\sum_{j\neq0} \langle s_0s_j\rangle_c\leq 2^d-1$ and hence there  exists $\beta_c <\infty$ such that $\frac{\partial}{\partial h}f_\beta(h)\big\vert_{h=0}=1$ and the paramagnetic solution $h=0$ becomes unstable when $\beta=\beta_c$. If the function $f_\beta(h)$ is convex in $h$ then for $\beta > \beta_c$ there is a unique (up to $h\rightarrow-h$) stable solution $\vert h\vert\neq0$ of the cavity field equation (\ref{def:f(h)}) which corresponds to the  ferromagnetic $m\neq0$ phase.  For $d=2$ this solution for $\beta > \beta_c=-\frac{1}{4}\log(\sqrt{5} - 2   )$ is given by 
\begin{eqnarray}
h=  \log\!\left(  \!\! \frac{   \sqrt{  \rme^{8\beta}   \!-\!   2\rme^{4\beta}  \! - \!  1  \!+ \! \sqrt{ \rme^{16\beta}   -    4\rme^{12\beta} \! -\!    2\rme^{8\beta}    \!+\!   4\rme^{4\beta}    \!+\!    1   }}}{
 \sqrt{2}\,\rme^{2\beta}}\!\right) \label{eq:h-2d},
\end{eqnarray}
but for $d=3,4$ we were able to obtain these solutions only numerically.

It follows from the equation (\ref{eq:C-hyper})  that in the paramagnetic phase ($h=0$) the specific heat is given by  the equation 
\begin{eqnarray}
C(\beta)&=&  2^{1-d}\beta^2\left[\langle E_0^2(s) \rangle -\langle E_0(s) \rangle^2\right].\label{eq:C-hyper-high-T}
\end{eqnarray}
and that it is bounded on the interval $\beta\in[0, \beta_c]$. For $\beta > \beta_c$ the system is in the ferromagnetic phase ($h\neq0 $) and the specific heat is given by the equation 

\begin{eqnarray}
C(\beta)&=&2^{1-d}\beta^2  \Big\{\left[\langle E_0^2(s) \rangle -\langle E_0(s) \rangle^2\right] \label{eq:C-hyper-low-T}\\
&&-  \sum_{j}\left[\langle E_0(s) s_j\rangle -\langle E_0(s) \rangle \langle s_j \rangle\right]\frac{\partial h}{\partial\beta}\Big\}.\nonumber
\end{eqnarray}

Let us consider the last line of the  above equation. The correlation $-\left[\langle E_0(s) s_j\rangle -\langle E_0(s) \rangle \langle s_j \rangle\right]=\sum_A J_A \left[\langle s_A s_j\rangle -\langle s_A \rangle \langle s_j \rangle\right]$  is bounded and positive (by the GKS theorem~\cite{Domb1972}). Furthermore,  it follows from the magnetization equation (\ref{eq:m(h)})  that the derivative $\frac{\partial h}{\partial\beta}= \frac{\partial m}{\partial\beta}\frac{1}{ 2(1-m^2)}$ is also positive ($m$ is a monotonic non-decreasing function of $\beta$) but is not necessary bounded.  For $d=2$ it diverges as $\beta\rightarrow\beta_c^{+}$,  but the $\lim_{\beta\rightarrow\beta_c^{+}} C(\beta)=\log^2(\sqrt{5}  -2 )(1956577837718\sqrt{5}-4375041048407)/(1182573459758\sqrt{5} - 2644314644406)\approx2.283$ is bounded and as $\beta\rightarrow\beta_c^{-}$  the $\lim_{\beta\rightarrow\beta_c^{-}} C(\beta)=\log^2(\sqrt{5}  -2 )(   1203    -538\sqrt{5}  )/(   110\sqrt{5}  - 246   )\approx 0.293$,  so the specific heat has a jump (see Figure  \ref{figure:2d}) at $\beta_c=-\frac{1}{4}\log(\sqrt{5} - 2 )$.  The specific heat $C(\beta)$ is bounded for any finite $d$ as $\beta\rightarrow\beta_c^{-}$ (this follows from the equation (\ref{eq:C-hyper-high-T})), but for $\beta\rightarrow\beta_c^{+}$  we were able to verify this behaviour (as in Figure  \ref{figure:2d})  only numerically in $d=3,4$.

\section{Formulae for Ising models on hypercubic lattices with $d=4$}
\label{app:formulas}

For $d=4$ the formulae for the cavity field $h$, the magnetisation per spin $m$ and the energy density $E$ become, respectively, 
\begin{eqnarray}
h&=&\frac{1}{2}\ln\big[\big(
264 {\rme^{-5 h+8 \beta}}+64 {\rme^{-9 h+
12 \beta}}+60 {\rme^{12 \beta-7 h}}+39 {\rme^{8 
\beta-11 h}}\label{eq:4d-cavity}\\
&&+184 {\rme^{8 \beta-9 h}}+215 {
\rme^{8 \beta-7 h}}+36 {\rme^{12 \beta-5 h}}+520 {\rme^{4 \beta-7 h}}\nonumber\\
&&+{\rme^{24 \beta-15 h}}+4 {\rme^{20 \beta-13 h}}+11 {\rme^{-13
 h+16 \beta}}+18 {\rme^{16 \beta-11 h}}\nonumber\\
&&+6 {
\rme^{16 \beta-9 h}}+48 {\rme^{12 \beta-11 h}}+28 {\rme^{12 \beta-3 h}}+57 {\rme^{-9 h}}+4 {\rme^{16 \beta-h}}\nonumber\\
&&+144 {\rme^{4 
\beta-9 h}}+684 {\rme^{4 \beta-5 h}}+273 {
\rme^{8 \beta-3 h}}+1098 {\rme^{-5 h}}\nonumber\\
&&+840
 {\rme^{-3 h+4 \beta}}+384 {\rme^{8 \beta-h}}+140 {\rme^{-4 \beta-7 h}}+36 {\rme^{12 
\beta+h}}\nonumber\\
&&+390 {\rme^{-7 h}}+1610 {\rme^{-3
 h}}+40 {\rme^{-8 \beta-7 h}}+588 {\rme^{
-4 \beta-5 h}}\nonumber\\
&&+800 {\rme^{4 \beta-h}}+2178 {
\rme^{-h}}+351 {\rme^{8 \beta+h}}+288 {
\rme^{-8 \beta-5 h}}\nonumber\\
&&+1372 {\rme^{-4 \beta-3 h}}+1080 {\rme^{4 \beta+h}}+60 {\rme^{12 \beta+
3 h}}+609 {\rme^{-8 \beta-3 h}}\nonumber\\
&&+1632 {
\rme^{-h-4 \beta}}+2070 {\rme^{h}}+440 {
\rme^{8 \beta+3 h}}+1140 {\rme^{4 \beta+3 h}}\nonumber\\
&&+132 {\rme^{12 \beta+5 h}}+21 {\rme^{-16 
\beta-5 h}}+224 {\rme^{-12 \beta-3 h}}\nonumber\\
&&+1056 
{\rme^{-h-8 \beta}}+1764 {\rme^{h-4 \beta}}
+1830 {\rme^{3 h}}\nonumber\\
&&+473 {\rme^{8 \beta+5 h
}}+18 {\rme^{16 \beta+7 h}}+24 {\rme^{-12 \beta-5
 h}}+783 {\rme^{h-8 \beta}}\nonumber\\
&&+288 {\rme^{-12 \beta-h}}+980 {
\rme^{3 h-4 \beta}}+1144 {\rme^{4 \beta+5 h}}+192 {\rme^{12 \beta+7 h}}\nonumber\\
&&+7 {\rme^{-24 
\beta-3 h}}+60 {\rme^{-16 \beta-h}}+32 {
\rme^{-20 \beta-h}}+288 {\rme^{h-12 \beta}}
\nonumber\\
&&+480 {\rme^{3 h-8
 \beta}}+858 {\rme^{5 h}}+552 {\rme^{8 \beta+7 h}}+78 {\rme^{16 \beta+9 h}}\nonumber\\
&&+308 {\rme^{5 h-4 \beta}}+432 {
\rme^{4 \beta+7 h}}+208 {\rme^{12 \beta+9 h}}+40 {\rme^{-12 \beta+3 h}}\nonumber\\
&&+28 {\rme^{20 
\beta+11 h}}+{\rme^{-32 \beta-h}}+9 {\rme^{
-24 \beta+h}}+35 {\rme^{3 h-16 \beta}}\nonumber\\
&&+88 {
\rme^{-8 \beta+5 h}}+42 {\rme^{-16
 \beta-3 h}}+54 {\rme^{-16 \beta+h}}+{\rme^{32 
\beta+15 h}}\nonumber\\
&&+171 {\rme^{7 h}}+169
 {\rme^{8 \beta+9 h}}+77 {\rme^{16 \beta+11 h}}+15 {\rme^{24 \beta+13 h}}\big)\nonumber\\
&&\times
\big( 
473 {\rme^{-5 h+8 \beta}}+208 {\rme^{-9 h
+12 \beta}}+192 {\rme^{12 \beta-7 h}}+169 {\rme^{8
 \beta-9 h}}\nonumber\\
&&~~+552 {\rme^{8 \beta-7 h}}+132 {
\rme^{12 \beta-5 h}}+432 {\rme^{4 \beta-7 h}}+77 {\rme^{16 \beta-11 h}}\nonumber\\
&&~~+78 {\rme^{16 
\beta-9 h}}+60 {\rme^{12 \beta-3 h}}+351 {\rme^{8 \beta-h}}+1080 {\rme^{4 \beta-h}}\nonumber\\
&&~~+1144 {
\rme^{4 \beta-5 h}}+440 {\rme^{8 \beta-3 {\it h
}}}+858 {\rme^{-5 h}}+1140 {\rme^{-3 h+4 
\beta}}\nonumber\\
&&~~+171 {\rme^{-7 h}}+1830 {\rme^{-3 h}}+308 {\rme^{-4 \beta-5 h}}\nonumber\\
&&~~+2070 {\rme^{-h}}+384 {\rme^{8 \beta+h}}+88 {\rme^{-8 
\beta-5 h}}+980 {\rme^{-4 \beta-3 h}}\nonumber\\
&&~~+800 {
\rme^{4 \beta+h}}+273 {\rme^{8 \beta+3h}}+1056 {\rme^{h-8 \beta}}\nonumber\\
&&~~+28 {\rme^{12 \beta+3 h}}
+480 {\rme^{-8 \beta-3 h}}+1764 {\rme^{-h-
4 \beta}}+2178 {\rme^{h}}\nonumber\\
&&~~+840 {\rme^{4 \beta+3 h}}+36 {\rme^{12 
\beta+5 h}}+40 {\rme^{-12 \beta-3 h}}\nonumber\\
&&~~+783 {
\rme^{-h-8 \beta}}+1632 {\rme^{h-4 \beta}}+
1610 {\rme^{3 h}}+264 {\rme^{8 \beta+5 h}
}\nonumber\\
&&~~+35 {\rme^{-16 \beta-3 h}}+288 {\rme^{-12 \beta-h}}+1372 {\rme^{3 
h-4 \beta}}+684 {\rme^{4 \beta+5 h}}\nonumber\\
&&~~+60 {
\rme^{12 \beta+7 h}}+54 {\rme^{-16 \beta-h
}}+288 {\rme^{h-12 \beta}}\nonumber\\
&&~~+60 {\rme^{-16 \beta+h}}+609 {\rme^{3 h-8 \beta}}+1098 {\rme^{5
 h}}+215 {\rme^{8 \beta+7 h}}\nonumber\\
&&~~+6 {\rme^{16
 \beta+9 h}}+18 {
\rme^{16 \beta+11 h}}+4 {
\rme^{16 \beta+h}}+24 {\rme^{-12 \beta+5 h}}\nonumber\\
&&~~+588 {\rme^{5 h-4 \beta}}+520 {
\rme^{4 \beta+7 h}}+64 {\rme^{12 \beta+9 h}}+224 {\rme^{-12 \beta+3 h}}\nonumber\\
&&~~+42 {\rme^{3 h-16 \beta}}+288 {\rme^{-8 \beta+5 h}}+390 {
\rme^{7 h}}+184 {\rme^{8 \beta+9 h}}\nonumber\\
&&~~+9 {\rme^{-24 \beta-h
}}+32 {\rme^{-20 \beta+h}}+{\rme^{32 \beta-15 h}}+15 {\rme^{24 \beta-13 h}}\nonumber\\
&&~~+28 {\rme^{20 
\beta-11 h}}+40 {\rme^{7 h-8 \beta}}+57 {
\rme^{9 h}}+39 {\rme^{8 \beta+11 h}}\nonumber\\
&&~~+36 {\rme^{12 \beta-h}}+
11 {\rme^{16 \beta+13 h}}+{\rme^{-32 \beta+h}}+7 {\rme^{-24 \beta+3 h}}\nonumber\\
&&~~+21 {\rme^{5 h-16 \beta}}+48 {\rme^{12 \beta+11 h}}+18 {
\rme^{16 \beta-7 h}}\nonumber\\
&&~~+140 {\rme^{7 h-4 \beta}}+144 {\rme^{4 
\beta+9 h}}+4 {\rme^{20 \beta+13 h}}+{\rme
^{24 \beta+15 h}}
\big)^{-1}\big]\nonumber
\\[2mm]
m&=&%
\big(-12 {\rme^{-16 \beta-2 h}}+12 {\rme^{16 \beta+8 h}}-16 {\rme^{-12 \beta-4 h}}+60 {\rme^{16
 \beta+10 h}}\label{eq:4d-m}\\
 &&+174 {\rme^{2 h-8 \beta}}+24 {\rme^{20 \beta+12 h}}+732 {\rme^{4 h}}+258{\rme^{8 \beta+6 h}}\nonumber\\
&&
 +48 {\rme^{-8 \beta+6 h}}+2 {\rme^{-24 \beta+2 h}}+14 {\rme^{4 h-16 \beta}}+114 {\rme^{8 h}}+130 {\rme^{8 \beta+10 h}}\nonumber\\
&&+66 {\rme^{16 \beta+12 h}}+392 {\rme^{4 h-4 \beta}}+624 {\rme^{4 \beta+6 h}}+128 {\rme^{12 \beta+8 h}}\nonumber\\
&&+12 {\rme^{-16 \beta+2 h}}+192 {\rme^{4 h-8 \beta}}+468 {\rme^{6 h}}+64 {\rme^{2 h-12 \beta}}\nonumber\\
&&+14 {\rme^{24 \beta+14 h}}+168 {\rme^{6 h-4 \beta}}+16{\rme^{-12 \beta+4 h}}\nonumber\\
&&+288 {\rme^{4 \beta+8 h}}+160 {\rme^{12 \beta+10 h}}+{\rme^{32 \beta+16 h}}\nonumber\\
&&-{\rme^{32 \beta-16 h}}-14 {\rme^{24 \beta-14 h}}-24 {\rme^{20 \beta-12 h}}\nonumber\\
&&-66 {\rme^{-12 h+16 \beta}}-60 {\rme^{16 \beta-10 h}}-12 {\rme^{16 \beta-8 h}}\nonumber\\
&&-160 {\rme^{12 \beta-10 h}}-128 {\rme^{12 \beta-8 h}}-176 {\rme^{-4 h+8 \beta}}\nonumber\\
&&-8 {\rme^{12 \beta-2 h}}-130 {\rme^{8 \beta-10 h}}-368 {\rme^{8 \beta-8 h}}\nonumber\\
&&-72 {\rme^{12 \beta-6 }}-258 {\rme^{8 \beta-6 h}}-24 {\rme^{12 \beta-4h}}\nonumber\\
&&-114 {\rme^{-8 h}}-624 {\rme^{4 \beta-6 h}}-288 {\rme^{4 \beta-8 h}}-468 {\rme^{-6 h}}\nonumber\\
&&-456 {\rme^{4 \beta-4 h}}-732 {\rme^{-4 h}}-78 {\rme^{8 \beta-2 h}}-48 {\rme^{-8 \beta-6 h}}\nonumber\\
&&-392 {\rme^{-4 \beta-4 h}}-240 {\rme^{-2 h+4 \beta}}-460 {\rme^{-2 h}}-168 {\rme^{-4 \beta-6 h}}\nonumber\\
&&+8 {\rme^{12 \beta+2 h}}-192 {\rme^{-8 \beta-4 h}}-392 {\rme^{-4 \beta-2 h}}+78 {\rme^{8 \beta+2 h}}\nonumber\\
&&+240 {\rme^{4 \beta+2 h}}+24 {\rme^{12 \beta+4h}}-14 {\rme^{-16 \beta-4 h}}-174 {\rme^{-8 \beta-2 h}}\nonumber\\
&&-64 {\rme^{-12 \beta-2 h}}+460 {\rme^{2 h}}+176 {\rme^{8 \beta+4 h}}+392 {\rme^{2 h-4 \beta}}\nonumber\\
&&+456 {\rme^{4 \beta+4 h}}+72 {\rme^{12 \beta+6 h}}-2 {\rme^{-24 \beta-2 h}}+368 {\rme^{8 \beta+8 h}}\big)/\mathcal{N}\nonumber
\\[2mm]
E&=&
-4\big(12 {\rme^{24 \beta+14 h}}+{\rme^{32 \beta+16 {  
h}}}+{\rme^{32 \beta-16 h}}+12 {\rme^{24 \beta-
14 h}}\label{eq:4d-E2}\\
&&~~+20 {\rme^{20 \beta-12 h}}+44 {\rme^{-12 h+16 \beta}}+48 {\rme^{16 \beta-10 h}}+12 {\rme^{16 \beta-8 h}}\nonumber\\
&&~~+96 {\rme^{12 \beta-10 {h}}}+96 {\rme^{12 \beta-8 h}}+52 {\rme^{8 \beta-10 h}}+184 {\rme^{8 \beta-8 h}}\nonumber\\
&&~~+72 {\rme^{12 \beta-6 h}}+172 {\rme^{8 \beta-6 {  h}}}+36 {\rme^{12 \beta-4 h}}+208 {\rme^{4 \beta-6 h}}\nonumber\\
&&~~+176 {\rme^{-4 h+8 \beta}}+24 {\rme^{12 \beta-2 h}}+4 {\rme^{16 \beta}}+72 {\rme^{4 \beta-8 h}}\nonumber\\
&&~~+228 {\rme^{4 \beta-4 {  h}}}+156 {\rme^{8 \beta-2 h}}-32 {\rme^{-8 \beta-6h}}-196 {\rme^{-4 \beta-4 h}}\nonumber\\
&&~~+240 {\rme^{-2 h+4 \beta}}+192 {\rme^{8 \beta}}-56 {\rme^{-4\beta-6 h}}+200 {\rme^{4 \beta}}\nonumber\\
&&~~+24 {\rme^{12 \beta+2 h}}-192 {\rme^{-8 \beta-4 h}}-392 {\rme^{-4 \beta-2 h}}+156 {\rme^{8 \beta+2 {  h}}}\nonumber\\
&&~~+240 {\rme^{4 \beta+2 h}}+36 {\rme^{12 \beta+4 h}}-28 {\rme^{-16 \beta-4 h}}-348 {\rme^{-8 \beta-2 h}}\nonumber\\
&&~~-192 {\rme^{-12 \beta-2 {  h}}}-408 {\rme^{-4 \beta}}-528 {\rme^{-8 \beta}}+176 {\rme^{8 \beta+4 h}}\nonumber\\
&&~~-392 {\rme^{2 h-4 \beta}}+228 {\rme^{4 \beta+4 h}}+72 {\rme^{12 \beta+6 h}}+172 {\rme^{8 \beta+6 h}}\nonumber\\
&&~~+12 {\rme^{16 \beta+8 h}}-24 {\rme^{-12 \beta-4 {  h}}}-48 {\rme^{-16 \beta-2 h}}-216 {\rme^{-12 \beta}}\nonumber\\
&&~~-348 {\rme^{2 h-8 \beta}}-196 {\rme^{4 {h}-4 \beta}}+208 {\rme^{4 \beta+6 h}}+96 {\rme^{12 \beta+8 h}}\nonumber\\
&&~~-12 {\rme^{-24 \beta-2 {  h}}}-60 {\rme^{-16 \beta}}-40 {\rme^{-20 \beta}}-192 {\rme^{2 h-12 \beta}}\nonumber\\
&&~~-48 {\rme^{-16 \beta+2 {  h}}}-192 {\rme^{4 h-8 \beta}}+184 {\rme^{8 \beta+8 h}}+48 {\rme^{16 \beta+10 h}}\nonumber\\
&&~~-56 {\rme^{6 h-4 \beta}}-24 {\rme^{-12 \beta+4 {  h}}}+72 {\rme^{4 \beta+8 h}}+96 {\rme^{12 \beta+10 h}}\nonumber\\
&&~~+20 {\rme^{20 \beta+12 h}}-2 {\rme^{-32 \beta}}-12 {\rme^{-24 \beta+2 h}}-28 {\rme^{4 h-16 \beta}}\nonumber\\
&&~~-32 {\rme^{-8 \beta+6 h}}+52 {\rme^{8 \beta+10 h}}+44 {\rme^{16 \beta+12 {  h}}}\big)/\mathcal{N}\nonumber
\end{eqnarray}
with the normalisation factor
\begin{eqnarray}
\mathcal{N}&=&
4356+96 {\rme^{-16 \beta-2 {  h}}}+24 {\rme^{16 \beta+8 {  h}}}+64 {\rme^{-12 \beta-4 {  h}}}\label{eq:4d-Norm}\\
&&
+1392 {\rme^{2 {  h}-8 \beta}}+32 {\rme^{20 \beta+12 {  h}}}+2928 {\rme^{4 {  h}}}+688 {\rme^{8 \beta+6 {  h}}}\nonumber\\
&&+128 {\rme^{-8 \beta+6{  h}}}+16 {\rme^{-24 \beta+2 {  h}}}+56 {\rme^{4 {  h}-16 \beta}}+228 {\rme^{8 {  h}}}\nonumber\\
&&+208 {\rme^{8 \beta+10 {  h}}}+88 {\rme^{16 \beta+12 {  h}}}+1568 {\rme^{4 {  h}-4 \beta}}+1664 {\rme^{4 \beta+6 {h}}}\nonumber\\
&&+256 {\rme^{12 \beta+8 {  h}}}+96 {\rme^{-16\beta+2 {  h}}}+768 {\rme^{4 {  h}-8 \beta}}+1248 {\rme^{6 {  h}}}\nonumber\\
&&+512 {\rme^{2 {  h}-12 \beta}}+16{\rme^{24 \beta+14 {  h}}}+448 {\rme^{6 {  h}-4\beta}}+64 {\rme^{-12 \beta+4 {  h}}}\nonumber\\
&&+576 {\rme^{4 \beta+8 {  h}}}+256 {\rme^{12 \beta+10 {  h}}}+{\rme^{32 \beta+16 {  h}}}+{\rme^{32 \beta-16 {  h}}}\nonumber\\
&&+16 {\rme^{24 \beta-14 {  h}}}+32 {\rme^{20 \beta-12{  h}}}+88 {\rme^{-12 {  h}+16 \beta}}+96 {\rme^{16 \beta-10 {  h}}}\nonumber\\
&&+24 {\rme^{16 \beta-8 {  h}}}+256 {\rme^{12 \beta-10 {  h}}}+256 {\rme^{12 \beta-8 {  h}}}+704 {\rme^{-4 {  h}+8 \beta}}\nonumber\\
&&+64 {\rme^{12\beta-2 {  h}}}+208 {\rme^{8 \beta-10 {  h}}}+736 {\rme^{8 \beta-8 {  h}}}+192 {\rme^{12 \beta-6 {  h}}}\nonumber\\
&&+688 {\rme^{8 \beta-6 {  h}}}+96 {\rme^{12 \beta-4 {  h}}}+228 {\rme^{-8 {  h}}}+1664 {\rme^{4 \beta-6 {  h}}}\nonumber\\
&&+576 {\rme^{4 \beta-8 {  h}}}+1248{\rme^{-6 {  h}}}+1824 {\rme^{4 \beta-4 {  h}}}+2928 {\rme^{-4 {  h}}}\nonumber\\
&&+624 {\rme^{8 \beta-2 {  h}}}+128 {\rme^{-8 \beta-6 {  h}}}+1568 {\rme^{-4 \beta-4 {  h}}}+1920 {\rme^{-2 {  h}+4 \beta}}\nonumber\\
&&+3680 {\rme^{-2 {  h}}}+448 {\rme^{-4 \beta-6 {  h}}}+64{\rme^{12 \beta+2 {  h}}}+768 {\rme^{-8 \beta-4 { h}}}\nonumber\\
&&+3136 {\rme^{-4 \beta-2 {  h}}}+624 {\rme^{8\beta+2 {  h}}}+1920 {\rme^{4 \beta+2 {  h}}}+96 {\rme^{12 \beta+4 {  h}}}\nonumber\\
&&+56 {\rme^{-16 \beta-4 {  h}}}+1392 {\rme^{-8 \beta-2 {  h}}}+512 {\rme^{-12 \beta-2 {  h}}}+3680 {\rme^{2 {  h}}}\nonumber\\
&&+704 {\rme^{8\beta+4 {  h}}}+3136 {\rme^{2 {  h}-4 \beta}}+1824{\rme^{4 \beta+4 {  h}}}+192 {\rme^{12 \beta+6 {  h}}}\nonumber\\
&&+16 {\rme^{-24 \beta-2 {  h}}}+736 {\rme^{8 \beta+8 {  h}}}+576 {\rme^{-12 \beta}}+2 {\rme^{-32 \beta}}\nonumber\\
&&+64 {\rme^{-20 \beta}}+8 {\rme^{16 \beta}}+768 {\rme^{8 \beta}}+1600 {\rme^{4 \beta}}+3264 {\rme^{-4 \beta}}\nonumber\\
&&+2112 {\rme^{-8 \beta}}+120 {\rme^{-16 \beta}}+96 {\rme^{16 \beta+10 {  h}}}
\nonumber
\end{eqnarray}

\end{document}